\documentclass[aps,prc,twocolumn,showpacs,floatfix,amsmath,amssymb,nofootinbib,longbibliography]{revtex4-1}

\usepackage{graphicx}
\usepackage{dcolumn}
\usepackage{hyperref}
\usepackage{bm}
\usepackage{multirow}
\usepackage{nicefrac}

\newcolumntype{d}[1]{D{.}{.}{#1}}

\graphicspath{{.}{./fig-article/}}

\newcommand{\nm}{\ensuremath{N_\mathrm{max}}}
\newcommand{\nmsp}{\ensuremath{N_\mathrm{sp,max}}}
\newcommand{\ho}{\ensuremath{\hbar \omega}}

\newcommand{\be}{\begin{equation}}
\newcommand{\ee}{\end{equation}}
\newcommand{\ba}{\begin{eqnarray}}
\newcommand{\ea}{\end{eqnarray}}

  \def\nuc#1#2{\relax\ifmmode{}^{#1}{\protect\text{#2}}\else${}^{#1}$#2\fi}
  \def\itnuc#1#2{\setbox\@tempboxa=\hbox{\scriptsize\it #1}
    \def\@tempa{{}^{\box\@tempboxa}\!\protect\text{\it #2}}\relax
    \ifmmode \@tempa \else $\@tempa$\fi}

\begin{document}

\title{Large-scale exact diagonalizations reveal low-momentum scales of nuclei}

\author{C.~Forss\'en} \email{christian.forssen@chalmers.se}
\affiliation{Department of Physics,
  Chalmers University of Technology, SE-412 96 G\"oteborg, Sweden}

\author{B.~D.~Carlsson} \affiliation{Department of Physics,
  Chalmers University of Technology, SE-412 96 G\"oteborg, Sweden}

\author{H.~T.~Johansson}
\affiliation{Department of Physics,
  Chalmers University of Technology, SE-412 96 G\"oteborg, Sweden}

\author{D.~S\"a\"af}
\affiliation{Department of Physics,
  Chalmers University of Technology, SE-412 96 G\"oteborg, Sweden}

\author{A.~Bansal} 
\affiliation{Department of Physics and
  Astronomy, University of Tennessee, Knoxville, Tennessee 37996, USA}
\affiliation{Physics Division, Oak Ridge National Laboratory, Oak
  Ridge, Tennessee 37831, USA}

\author{G.~Hagen}\thanks{This manuscript has been co-authored by
  UT-Battelle, LLC under Contract No. DE-AC05-00OR22725 with the
  U.S. Department of Energy. The United States Government retains and
  the publisher, by accepting the article for publication,
  acknowledges that the United States Government retains a
  non-exclusive, paid-up, irrevocable, world-wide license to publish
  or reproduce the published form of this manuscript, or allow others
  to do so, for United States Government purposes. The Department of
  Energy will provide public access to these results of federally
  sponsored research in accordance with the DOE Public Access
  Plan. (http://energy.gov/downloads/doe-public-access-plan).}
\affiliation{Department of Physics and
  Astronomy, University of Tennessee, Knoxville, Tennessee 37996, USA}
\affiliation{Physics Division, Oak Ridge National Laboratory, Oak
  Ridge, Tennessee 37831, USA}

\author{T.~Papenbrock} 
\affiliation{Department of Physics and
  Astronomy, University of Tennessee, Knoxville, Tennessee 37996, USA}
\affiliation{Physics Division, Oak Ridge National Laboratory, Oak
  Ridge, Tennessee 37831, USA}

\begin{abstract}
  \emph{Ab initio} methods aim to solve the nuclear many-body problem
  with controlled approximations. Virtually exact numerical solutions for
  realistic interactions can only be obtained for certain special
  cases such as few-nucleon systems.
  Here we extend the reach of exact diagonalization methods to
  handle model spaces with dimension exceeding $10^{10}$ on a single
  compute node.
  This allows us to perform no-core
  shell model (NCSM) calculations for
  \nuc{6}{Li} in model spaces up to $\nm = 22$ and to reveal the
  \nuc{4}{He}+d halo structure of this nucleus.
  Still, the use of a finite harmonic-oscillator basis implies
  truncations in both infrared (IR) and ultraviolet (UV) length
  scales. These truncations impose finite-size corrections on
  observables computed in this basis. We perform IR extrapolations of
  energies and radii computed in the NCSM and with the coupled-cluster
  method at several fixed UV cutoffs. It is shown that this strategy
  enables information gain also from data that is not
  fully UV converged.
  IR extrapolations improve the accuracy of relevant bound-state
  observables for a range of UV cutoffs, thus making them profitable
  tools.  We relate the momentum scale that governs the exponential IR
  convergence to the threshold energy for the first open decay
  channel.
  Using large-scale NCSM calculations we  numerically verify
  this small-momentum scale of finite nuclei.
 \end{abstract}


\maketitle

\section{Introduction}
The harmonic-oscillator basis continues to be most popular in the
computation of atomic nuclei. It is employed in the traditional shell
model~\cite{brown1988,caurier2005,shimizu2012}, nuclear density
functional calculations~\cite{schunck2012,stoitsov2013,niksic2014},
the no-core shell model (NCSM)~\cite{navratil2009,barrett2013}, and other
{\it ab initio} methods~\cite{dickhoff2004,hagen2014,hergert2016}.
For such {\it ab initio} approaches, one usually employs a very large
basis including many orbitals. This serves two simultaneous
purposes: (i) the basis should capture the correlations induced by the
strong, realistic nucleon-nucleon interactions that are used as input,
and (ii) it should build the long-range behavior of nuclear wave functions
and possibly incorporate effects of continuum coupling~\cite{forssen2013-2}. Still, the
basis must obviously be truncated and one might ask the relevant question:
What are the corrections to energies and other observables that are
due to the finite size of the oscillator basis?

This question was addressed in several works by empirical
extrapolation
schemes~\cite{horoi1999,zhan2004,hagen2007b,forssen2008,bogner2008}. Only
recently---based on the insight that a finite oscillator space
introduces IR and UV
cutoffs~\cite{stetcu2007,hagen2010b,jurgenson2011,coon2012}---were
extrapolation formulas derived for the harmonic-oscillator basis. The
IR extrapolation formulas~\cite{furnstahl2012} are the
harmonic-oscillator equivalent of L\"uscher's formula for the
lattice~\cite{luscher1985}.  The key insight was that the spherical
harmonic-oscillator basis is---at low energies---indistinguishable
from a spherical cavity of radius $L$. As the L\"uscher formula
corrects the energy shift from tunneling due to the periodic boundary
condition of the underlying lattice, the IR extrapolation formulas
correct the energy shift due to the Dirichlet boundary condition at the
radius $L$. Very recently, high-precision expressions for the length
scale $L$ were derived for the two-body problem~\cite{more2013},
many-body spaces that are products of single-particle
spaces~\cite{furnstahl2015}, and the NCSM~\cite{wendt2015} in which a
total energy truncation is employed. Extrapolation formulas were
derived for energies~\cite{furnstahl2012}, radii~\cite{more2013},
quadrupole transitions~\cite{odell2016}, and radiative capture
reactions~\cite{acharya2016}. For finite-volume corrections to the
binding energy of $N$-particle quantum bound states on the lattice we
refer the reader to Ref.~\cite{koenig2018}.

The leading-order (LO) IR extrapolation formula for energies is
\be
\label{IR}
E(L) = E_\infty + a_0\exp{(-2k_\infty L)}.
\ee
The energies $E(L)$ are theoretical results for bound-state energies,
while $a_0$ and $k_\infty$ are adjustable parameters that are so far
only understood in the two-body problem~\cite{more2013}. The IR
extrapolation~(\ref{IR}) reflects that a finite oscillator basis
effectively imposes a hard-wall boundary condition at a radius
$L$. Thus, $a_0> 0$ and the computed energies $E(L)$ are above the
infinite space result $E_\infty$. The extrapolation formula~(\ref{IR})
is expected to yield an accurate bound-state energy when UV
convergence is already achieved, and when $L$ significantly exceeds
any other relevant length scale, i.e., for $k_\infty L\to\infty$.  For
the deuteron, subleading corrections (in $k_\infty L$) to
Eq.~(\ref{IR}) are also known~\cite{furnstahl2014}.  We note that UV
extrapolations of bound-state energies are more challenging than IR
extrapolations because the former depend on the regulators and
short-range details of the employed interactions~\cite{koenig2014},
while the latter are insensitive to such details.

In practice, it is often challenging to fulfill the two conditions
(i.e., being both UV converged and working at asymptotically large values of
$k_\infty L$), and it would be profitable to relax these
conditions. We also note that IR extrapolations of bound-state
energies---when performed at large UV cutoffs that significantly
exceed the cutoff of the employed interaction---sometimes fail to
improve on the variational minimum; see
Refs.~\cite{furnstahl2015,wendt2015} for examples. This casts some
doubts on the usefulness of such extrapolations and makes it necessary
to revisit them in more detail. The development of a practical and
reliable scheme for IR extrapolations is a specific purpose of this
paper.

While the NCSM method promises many-body results without any
uncontrolled approximations, it often faces computational limits
in terms of both CPU and memory requirements. A second objective of
our work is to push the limit of the
exact-diagonalization method in nuclear physics. 
This extended reach will make it possible to probe how numerical
results depend on UV- and IR-scales. In particular, we will see
that the separation momentum of the lowest-lying decay channel is the
relevant low-momentum scale of bound states in finite nuclei.
An improved understanding of IR extrapolation can be employed
to optimize the choice of model-space parameters so that the
information yield of costly many-body calculations is maximized.

This paper is organized as follows: In Sec.~\ref{deriv} we propose
IR extrapolation formulas for energies and radii that are applicable
in cases lacking a full UV convergence. The extended reach of
large-scale exact diagonalization with the NCSM is presented in
Sec.~\ref{sec:pantoine}, with more details on the technical
developments that have made such calculations possible adjourned to
the Appendix \ref{sec:pantoine_details}.
We then present an extensive set of large-basis NCSM results and apply
the IR extrapolation formulas to several $s$- and $p$-shell nuclei in
Secs.~\ref{sec:sshell} and~\ref{sec:pshell}, respectively. We also
present results from coupled-cluster computations.
We summarize our results in Sect.~\ref{secsum}.

\section{Derivation}
\label{deriv}
Let us assume we work in model spaces with a fixed value of
$\Lambda$---the UV momentum cutoff scale---that is not yet so
large that UV convergence is fully achieved. Usually this is the case
for values of $\Lambda$ that only moderately exceed the cutoff
employed by the interaction. As the IR length $L$ is increased, the
tail of the bound-state wave function will be built up, and we see
that Eq.~(\ref{IR}) at fixed $\Lambda$ generalizes to
\be
\label{master}
E(L,\Lambda) = E_\infty(\Lambda) + a_0(\Lambda)\exp{\left[-2k_\infty(\Lambda) L\right]}.
\ee
Equation~(\ref{master}) is only the leading term for asymptotically
large $k_\infty L$ but exhibits the full $\Lambda$ dependence [at
  least for $\Lambda$ large enough to yield a bound-state energy
  $E(L,\Lambda)$]. We note that the combined IR and UV extrapolation
formula applied in Ref.~\cite{furnstahl2012} is a special case of
Eq.~(\ref{master}) with constant $k_\infty$,
  $a_0$ and $E_\infty(\Lambda) = E_\infty +
A_0\exp{(-2\Lambda^2/A_1^2)}$.

Let us discuss subleading corrections to Eq.~(\ref{master}).  Contributions of
partial waves with finite angular momentum lead to corrections
proportional to
\be
\label{sigma}
\sigma_\mathrm{IR}=\frac{\exp{[-2k_\infty(\Lambda)L]}}{k_\infty(\Lambda)L}.
\ee
Even smaller corrections are of order $\exp{(-4k_\infty L)}$.  So far,
little is known about corrections in nuclei consisting of three or
more nucleons. Below we will argue that $k_\infty$ is the momentum to
the first open separation channel (or particle-emission channel). In
nuclei with several open channels (e.g.\ separation of neutrons, of
protons, or of alpha particles), the leading corrections from each
channel are expected to be on the order of $\exp{[-2k_\mathrm{sep}(i)
  L]}$, where $k_\mathrm{sep}(i)$ is the separation momentum of
channel $i$. Such corrections could be sizable for particle emission
channels with similar energy thresholds and/or with sizable
asymptotic normalization coefficients (ANCs)~\cite{koenig2018,mukhamedzhanov1990}.

Let us consider applications of the extrapolation
formula~(\ref{master}) at fixed $\Lambda$. In the harmonic-oscillator
basis, the oscillator length is
\be
\label{bosc}
b \equiv \sqrt{\hbar\over m \omega}
\ee
for a nucleon mass $m$ and the oscillator frequency $\omega$. 
The IR length scale $L$ and the UV cutoff $\Lambda$ are related
to each other~\cite{koenig2014}
\ba
\label{L_Lambda}
L(N,b) &=& f(N)b,\nonumber \\
\Lambda(N,b) &=& f(N)\hbar b^{-1},
\ea
because of the complementarity of momenta and coordinates. Here, $f(N)$ is
a function that depends on the number $N$ of quanta that can be
excited. This function also depends on the number of particles and differs
for product spaces and NCSM spaces. We will use the standard notation
\nm{} to denote an NCSM truncation of \nm{} quanta above the
lowest possible configuration. The maximum number of quanta for a
single particle in such a basis will be, e.g., $N=\nm + 1$ for a
$p$-shell nucleus. Following \textcite{more2013},
$f(N)\approx[2(N+3/2+2)]^{1/2}$ when $N\gg 1$ for a two-body system in the
center-of-mass frame. In general, $f(N)\propto N^{1/2}$ for $N\gg
1$~\cite{furnstahl2014,wendt2015}. 

We can express $L$ in Eq.~(\ref{L_Lambda}) as $L(N,\Lambda)=\hbar
f^2(N)/\Lambda$. Thus, $L\propto N$ for $N\gg 1$ at fixed
$\Lambda$. This shows that IR extrapolations~(\ref{master}) at fixed
$\Lambda$ are actually exponential in $N$. Formally, this result
coincides with several commonly used extrapolation
formulas~\cite{horoi1999,zhan2004,hagen2007b,forssen2008,bogner2008,maris2009}. We
also note that this result agrees with semiclassical arguments
regarding the convergence of bound-states in the harmonic-oscillator
basis~\cite{littlejohn2002}.

For radii, we proceed as for the bound-state energies and generalize
the extrapolation formulas of Refs.~\cite{furnstahl2012,furnstahl2014}
to
\be
\label{radii}
r^2(L,\Lambda) = r^2_\infty(\Lambda) -
\alpha(\Lambda)\left[k_\infty(\Lambda) L\right]^3
\exp{\left[-2k_\infty(\Lambda) L\right]}
\ee
at fixed UV cutoff $\Lambda$. Here, 
corrections are of the size
\be
\label{sigmar}
\sigma_{r,\mathrm{IR}}=[k_\infty(\Lambda)L]\exp{[-2k_\infty(\Lambda)L]} 
\ee
for the two-body bound state.  As for the energies, there are other
radius corrections in nuclei consisting of three or more nucleons. For
these reasons, we will employ only the leading corrections,
i.e.\ Eq.~(\ref{master}) for the energies and Eq.~(\ref{radii}) for radii in
extrapolations of data. In the corresponding $\chi^2$ fits, we will
employ the uncertainties scaled with~(\ref{sigma}) and (\ref{sigmar}),
respectively.

In the extrapolation formulas~(\ref{master}) and (\ref{radii}), the
$\Lambda$-dependent quantities are taken as
adjustable parameters. In the deuteron, $k_\infty$ and $a_0$ are
related to the binding energy $B$ and the ANC via~\cite{furnstahl2014}
\be
\label{B}
B = {\hbar^2k_\infty^2\over 2\mu}, 
\ee

\be
\label{ANC_d}
\gamma_\infty^2 = \frac{\mu a_0}{\hbar^2 k_\infty}.
\ee
Here, $\mu=m/2$ is the reduced mass, $k_\infty$ is the separation
momentum, and $\gamma_\infty$ is the ANC defined by large-$r$ behavior
of the deuteron wave function in the relative coordinate
$\vec{r}=\vec{r}_1-\vec{r}_2$. We note that the oscillator length for
this coordinate employs the reduced mass instead of the nucleon mass
in Eq.~(\ref{bosc}). Below, we will employ length
and momentum scales that are based on Eq.~(\ref{bosc}).

We would like to understand the physics meaning of $k_\infty$ in IR
extrapolations of NCSM results for few- and many-body systems.  For
many-body bound states on a cubic lattice, this parameter was very
recently identified with the separation
momentum~\cite{koenig2018}. In what follows we arrive at a similar
identification for the harmonic-oscillator basis of the NCSM. In the
NCSM, the IR length~(\ref{L_Lambda}) constitutes an effective
hard-wall for the hyperradius $\rho$ with
\ba
\vec{\rho}\,^2 &=& \sum_{j=1}^A \vec{r}\,_i^2 - A \vec{R}\,_{\rm cm}^2 , 
\ea
where
\be
\vec{R}_{\rm cm} \equiv {1\over A} \sum_{j=1}^A \vec{r}_j
\ee
is the center of mass coordinate. We use an orthogonal transformation and
introduce Jacobi coordinates $\vec{\rho}_1,\ldots,\vec{\rho}_A$ such
that \ba
\label{jacobi}
\vec{\rho}_{A} &=&{A}^{1/2} \vec{R}_{\rm cm} = {1\over\sqrt{A}}\sum_{j=1}^A\vec{r}_j .
\ea
Using an orthogonal transformation has the advantage that the reduced
mass corresponding to each of the Jacobi coordinates is simply the
nucleon mass $m$. Thus, the oscillator length for each Jacobi
coordinate is given by Eq.~(\ref{bosc}).

\begin{figure}[tb]
\begin{center}
\includegraphics[width=0.95\columnwidth]{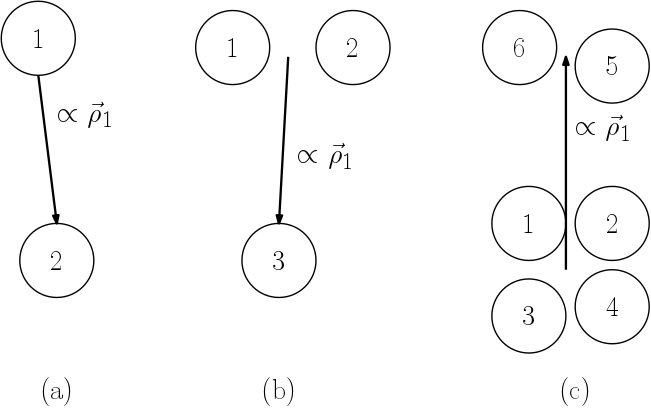}
\end{center}
\caption{Choice of Jacobi coordinates for the deuteron (a), the triton
  (b), and \nuc{6}{Li} (c) such that $\vec{\rho}_1$ corresponds to
the channel with the lowest separation energy.}
\label{fig:jacobi}
\end{figure}
We note that there are many ways to introduce Jacobi coordinates
$\vec{\rho}_1$ to $\vec{\rho}_{A-1}$ that are orthogonal to each other
and orthogonal to $\vec{\rho}_{A}$ in Eq.~\eqref{jacobi}. In
particular, one can choose $\vec{\rho}_1$ such that it corresponds to
the lowest-energetic separation channel.  See, for example, the
illustrations in Fig.~\ref{fig:jacobi}, where $\vec{\rho}_1 =
(\vec{r}_2-\vec{r}_1)/\sqrt{2}$ for the deuteron, $\vec{\rho}_1 =
[\vec{r}_3-(\vec{r}_1+\vec{r}_2)/2]\sqrt{2/3}$ for the triton (because
its lowest separation is into a neutron and a deuteron), and
$\vec{\rho}_1 = \left[ (\vec{r}_5+\vec{r}_6)/2 -
  (\vec{r}_1+\vec{r}_2+\vec{r}_3+\vec{r}_4)/4 \right]\sqrt{4/3}$ for
\nuc{6}{Li} (because its lowest separation threshold is into an alpha
particle and the deuteron). Here, we limit ourselves to breakup into
two clusters. For any orthogonal choice of Jacobi coordinates, the
intrinsic hyperradius is
\be
\rho^2 = \sum_{j=1}^{A-1} \rho_j^2.
\ee
We note that the effective hard-wall radius $L$ of the NCSM~\cite{wendt2015}
constitutes a hard-wall boundary condition also for the Jacobi
coordinate $\vec{\rho}_1$. Thus, bound-state wave functions in this
coordinate fall off asymptotically as $e^{-k_1 \rho_1}$, with
$\vec{k}_1$ being the momentum conjugate to $\vec{\rho}_1$. We denote
$k_1 \equiv k_\mathrm{sep}$ as the separation momentum, with the
corresponding separation energy,
\be
\label{eq:sep}
S = \frac{\hbar^2k_\mathrm{sep}^2}{2m},
\ee
where, $m$ is the nucleon mass. We note that this mass (opposed to a
reduced mass) enters here, because we used an orthogonal
transformation from $(\vec{r}_1,\ldots,\vec{r}_A)$ to the Jacobi
coordinates.  

As the IR extrapolation~\eqref{master} is based on the exponential
falloff $\sim\exp(-k_\infty \rho_1)$ of bound-state wave functions in
position space, we now identify
\be
\label{eq:kinf}
k_\infty=k_\mathrm{sep},
\ee
with the separation energy $S_\infty =
\hbar^2k_\infty^2 / (2m)$. Based on the 
derivation in Appendix~\ref{App_ANC} and Ref.~\cite{koenig2018}, we also
identify
\be
\label{eq:ANC}
a_0 = \frac{\hbar^2 k_{\rm sep} \gamma_{\rm sep}^2}{m},
\ee
where $\gamma_{\rm sep}$ is the ANC corresponding to the Jacobi coordinate
$\rho_1$. Taking the different choice of coordinates into account, we
note that Eqs.~\eqref{eq:sep}, \eqref{eq:kinf}, \eqref{eq:ANC} yield
the same value for the separation energy~(\ref{B})
in case of the two-body bound
state~\cite{furnstahl2014}.

We recall that the relation~\eqref{eq:kinf} between the momentum of
the lowest separation channel and the fit parameter $k_\infty$ from IR
extrapolations in the NCSM is valid only in the asymptotic regime
$k_\infty L\to\infty$. Many nuclei exhibit $n$ different separation
channels, with proton, neutron, and alpha-particle separation usually
being among the least energetic ones. These channels can correspond to
different orthogonal Jacobi coordinates or also to different choices
of Jacobi coordinates (that are not orthogonal to each other). In any
case, the corresponding momenta $k_1\le k_2\le \cdots\le k_n$ might
not be well separated in scale. In practical NCSM calculations one can
only reach the regime $k_1 L \gg 1$, and this means that other
separation channels can yield non-negligible corrections to the
leading-order IR extrapolation formulas~(\ref{master}) and
(\ref{radii}). In those cases, IR extrapolation will only yield an
approximate value for $k_\mathrm{sep}$, and the application of
Eqs.~(\ref{eq:sep}) and (\ref{eq:kinf}) will only yield an approximate
value for the separation energy.

In Secs.~\ref{sec:sshell} and~\ref{sec:pshell} we apply the
extrapolation formulas~(\ref{master}) and (\ref{radii}) to obtain
bound-state energies and radii at fixed $\Lambda$ for different $s$-
and $p$-shell nuclei, respectively. We use the nucleon-nucleon
interaction NNLO$_{\rm opt}$~\cite{ekstrom2013} with a regulator
cutoff $\Lambda_\chi=500$~MeV.
The nuclei \nuc{3}{H}, \nuc{3}{He}, and \nuc{4}{He} will serve as
examples where the IR extrapolation scheme and the interpretation of
the results can be validated by also performing converged NCSM
calculations. We will then study several $p$-shell systems:
\nuc{6}{Li}, \nuc{6,8}{He}, and \nuc{16}{O}. For \nuc{8}{He} we
benchmark IR extrapolated results at fixed $\Lambda$ from the NCSM and
the coupled-cluster method (CC)~\cite{hagen2014} while we use only the
CC method for \nuc{16}{O} .
%

\section{Exact diagonalization with the NCSM%
\label{sec:pantoine}}
The NCSM approach employed in this work has been described in several
papers; see, e.g., the reviews~\cite{navratil2009,barrett2013}. 
The main feature of this \emph{ab initio} method is the use of the harmonic-oscillator
basis, truncated by a chosen maximal total oscillator energy of the
$A$-nucleon system as defined by the model-space parameter \nm. The
Hamiltonian matrix is constructed in this basis and the relevant
eigensolutions are typically found using iterative diagonalization methods.

In the NCSM approach one does not make any approximations concerning
the structure of the many-body wave function. Therefore, the method
can, in principle, describe any kind of (bound) many-body state;
although the convergence might be slow in some cases, e.g., for systems
that exhibit a large degree of
clusterization or very low separation thresholds. The main
disadvantage of this method is the rapid growth in
model-space size with the number of particles and \nm\ truncation. In
many NCSM studies one employs
basis-dependent unitary transformations to speed up convergence. 
In this work, however, we use bare nuclear
interactions. 

We discuss the frontier of NCSM calculations in terms of
model-space dimension and matrix sizes in the next subsection,
before describing the NCSM code \texttt{pANTOINE} that has been used
in this work.

\subsection{Pushing the frontier of exact diagonalization%
\label{subsec:ncsmfrontier}}
\begin{figure}[tb]
\begin{center}
\includegraphics[width=\columnwidth]{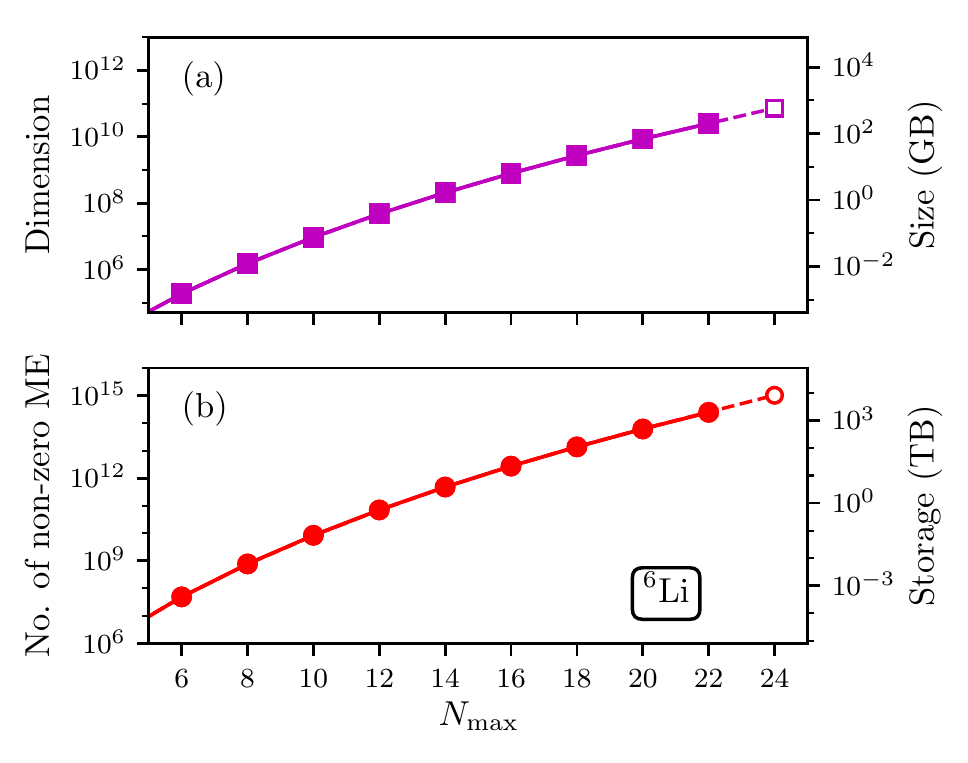}
\end{center}
\caption{Scaling plots for the \nuc{6}{Li} nuclear many-body problem
  ($M=1$) as a function of the NCSM model space truncation $\nm$. (a) Model space
  dimension; (b) Number of nonzero matrix elements (with two-body
  interactions only). The right axes displays the corresponding size (in TB)  assuming that
  the elements are explicitly stored in double-precision floating-point
  format. Extrapolated data is shown as open symbols. %
  \label{fig:ncsmscaling}}
\end{figure}

Let us use \nuc{6}{Li} as an example of NCSM dimensions and matrix
sizes. The M-scheme ($M=1$) model space dimension as a function of
\nm\ is shown on a semi-logarithmic scale in the upper panel of
Fig.~\ref{fig:ncsmscaling}. We note that 64-bit indices are needed 
when the dimension exceeds $4.2\cdot 10^9$ (for \nuc{6}{Li}
this occurs at $\nm=20$). For such dimensions it also becomes
difficult to fit the full vector in the machine memory.

However, the number of non-zero matrix elements, and the corresponding
number of operations needed for matrix-vector multiplications, is the most
restricting factor for these calculations. Restricting ourselves to
two-body interactions, the number of non-zero matrix elements for
\nuc{6}{Li} is shown in the lower panel of
Fig.~\ref{fig:ncsmscaling}. The data for this figure is generated employing the symmetry of the
Hamiltonian matrix, counting only matrix elements in the upper half of the
matrix.
A staggering amount of 2~PB memory storage space would be
required for the $\nm=22$ calculations assuming that we would explicitly store the matrix
using double precision.
In order to provide a relevant perspective on this number we note that
the most memory given for a machine on the current TOP500 list is 1.6 PB~\cite{top5002017}.
%
Obviously, the inclusion of three-body interaction terms would make this
problem even more dramatic.

Let us also comment on the efficiency limit for performing the
matrix-vector multiplications that are needed for iterative
diagonalization methods.  While an explicit-matrix code likely can
perform 1 multiplication (mult) per clock cycle in each core, a more restrictive
limit will be given by the memory bandwidth.  Assume the elements are
organized such that the vector data resides in processor cache and
thus memory bandwidth can be fully utilized to load indices and matrix
data.  Each element processed corresponds to about 10 bytes loaded, in
a streaming fashion.  State-of-the art server CPUs have eight channels of
DDR4 memory that can deliver just over 20 GB/s, and thus sustain 16
Gmult/s per socket.
At the same time, such server CPUs may have 32 cores running at 2.7
GHz, giving 86.4G clock cycles/s.  Memory bandwidth thus limits the
performed multiplications to 0.2 per clock cycle.  Utilizing the symmetry
of the Hamiltonian matrix, this becomes 0.4 mult/clock cycle. This
number constitutes an important performance benchmark for exact
diagonalization codes.

A possible solution to overcome the memory size limit is the implicit
(re-)construction of matrix elements rather than explicit storage. In
the next section we briefly discuss the sophisticated
shell-model and NCSM code \texttt{pANTOINE} that is designed to
achieve just that. We note that similar factorization techniques are used in the
configuration-interaction code \texttt{BIGSTICK}~\cite{Johnson:2013}.
Unavoidably, the bookkeeping that is needed to accomplish the
reconstruction of the matrix will lower the efficiency of the
calculations. We will show in Appendix~\ref{sec:pantoine_details} that
careful code design can limit this extra cost to about a factor four
in reduced efficiency.

\subsection{Exact diagonalization with implicit matrix construction%
\label{subsec:pantoine}}
\begin{figure}[tb]
\begin{center}
\includegraphics[width=0.8\columnwidth]{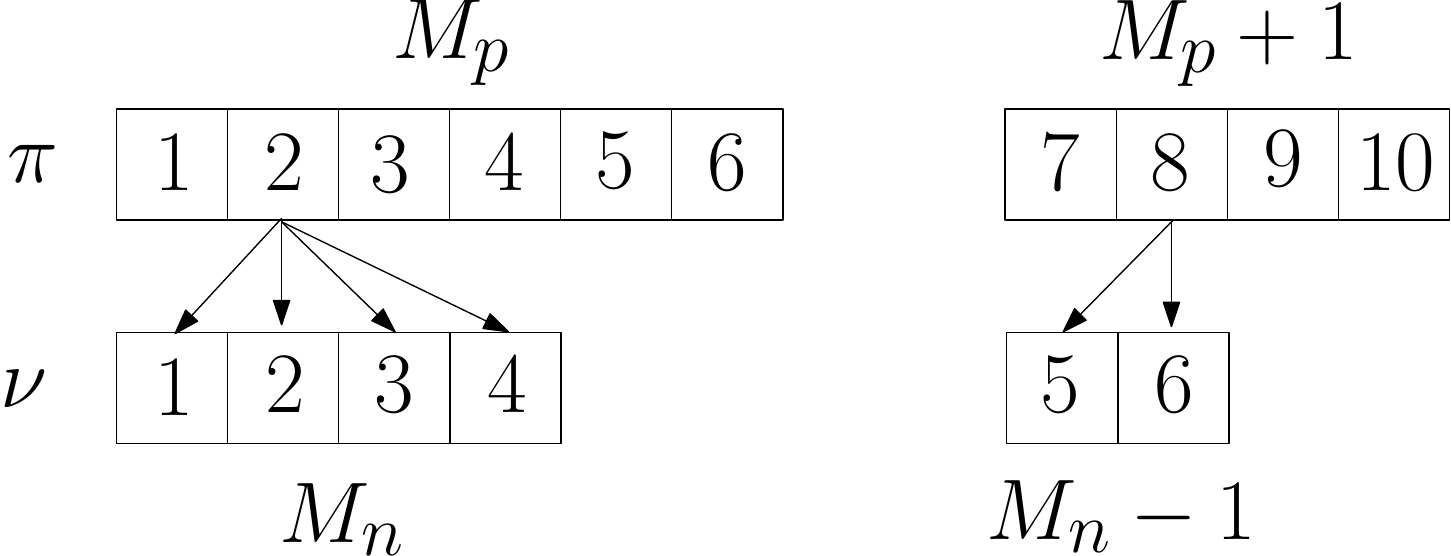}
\end{center}
\caption{Many-body states in the proton and neutron sub-spaces
  factorize into blocks according to their $J_z$
  projection. $A$-particle states with fixed $J_z=M$ are product
  states $| I \rangle = | \pi \rangle \otimes | \nu \rangle$ with
  proton- and neutron states from corresponding blocks.}
\label{fig:MpMn}
\end{figure}

The $A>4$ NCSM calculations presented in this work have been performed
with \texttt{pANTOINE}---an exact diagonalization code for nuclear
physics that is based on the NCSM version of \texttt{ANTOINE}
originally developed by Caurier and
coworkers~\cite{Caurier:1999tq,caurier1999,PhysRevC.69.014311}.
The main feature of this code is the implicit construction of the
Hamiltonian matrix, implying on-the-fly computation of matrix elements
in the iterative matrix-vector multiplications.  It employs the fact
that the total many-body space is a product of the much smaller spaces
spanned by protons and neutrons separately. A state $I$ in the full-space basis
can be labeled by a pair of proton ($\pi$) and neutron ($\nu$) states
in the subspace bases, as illustrated in Fig.~\ref{fig:MpMn}. All the
$\pi$ (and $\nu$) states are divided into blocks defined by their $J_z$
value.
To any proton block $J_{z,p} = M_p$ there is a corresponding neutron
block $J_{z,n} = M_n= M-M_p$, where $J_z=M$ is the total angular
momentum projection of the $A$-body state.
The full many-body basis is built by the association of proton states
$\pi$ (belonging to the block $M_p$) with neutron states $\nu$
(belonging to the corresponding neutron block $M_n$).
A simple numerical relation,
\begin{equation}
 I=R(\pi)+\nu,
\end{equation}
describing the index of a full multi-particle state can be
established.  Non-zero elements of the matrix, $H_{II'}=V(K)$, are
obtained through three integer additions: $I=R(\pi)+\nu$,
$I'=R(\pi')+\nu'$ and $K=Q(q_\pi)+q_\nu$. The index $q_\pi$ labels the
one-body operator acting between $\pi$ and $\pi'$ states, and
analogously the index $q_\nu$ links $\nu$ and $\nu'$ states.  The
storage of pre-calculated ($\pi$,$\pi'$,$q_\pi$) and
($\nu$,$\nu'$,$q_\nu$) labels remains possible as the dimensions in
respective proton- and neutron-spaces are moderate compared to those
of the full $A$-body space.  Note that each triple either applies
to the proton- or neutron-subspace only.  By performing a
double-loop over the pairs of such triple-lists, and performing the index
additions, all connections in the matrix can be efficiently processed.

With \texttt{pANTOINE} we have introduced several improvements of the
\texttt{ANTOINE} code and managed to significantly push the frontier
of exact diagonalization methods for few and many-nucleon systems. In
particular, we have achieved to extend six-body (\nuc{6}{Li}) NCSM calculations
with two-body interactions from $\nm=18$, which was the previous
computational limit~\cite{wendt2015,shin2017}, to $\nm=22$. This
translates to an increase of the model-space dimension by an order of
magnitude from $2.7 \cdot 10^{9}$ to $2.5 \cdot 10^{10}$.
More details on the technical development of our NCSM code can be
found in Appendix~\ref{sec:pantoine_details}. 

\section{$s$-shell nuclei: Validation and convergence
\label{sec:sshell}}
\subsection{$A=3$ nuclei
\label{subsec:A3}}

The three-nucleon bound states of \nuc{3}{H} and \nuc{3}{He} can be
computed virtually exactly. While there is little need for IR
extrapolations of these calculations, they allow us to validate the IR
extrapolation scheme and to check the relation~(\ref{eq:kinf}) between
the separation energy and the momentum $k_\infty$. The bound-state
energies of these nuclei are converged in the largest $\nm=40$ spaces
we employ, and we find $E(\nuc{3}{He})=-7.52$~MeV and
$E(\nuc{3}{H})=-8.25$~MeV. The corresponding separation energies with
respect to the deuteron [$E(\nuc{2}{H})=-2.2246$~MeV for the
  interaction NNLO$_{\rm opt}$] give $k_\mathrm{sep} \approx 0.50$ and
$0.54$~fm$^{-1}$ from Eq.~\eqref{eq:sep} for \nuc{3}{He} and
\nuc{3}{H}, respectively.

We fix the UV cutoff $\Lambda$, and for Hilbert spaces with $\nm\le
\mathrm{max}(\nm)$ compute the corresponding oscillator length
[i.e., the oscillator spacing $\ho(N,\Lambda)$], using the tables
presented in Ref.~\cite{wendt2015} for the function $f(N)$ in
Eq.~(\ref{L_Lambda}) for the nucleus \nuc{3}{H}.  This yields Hilbert
spaces with identical UV cutoffs and different IR lengths $L$.
At these fixed $\Lambda$, we compute the ground-state energies
$E(L,\Lambda)$ and point-proton radii $r(L,\Lambda)$ and perform IR
extrapolations.

\begin{figure}[hbtp]
  \centering
  \includegraphics[width=\columnwidth]{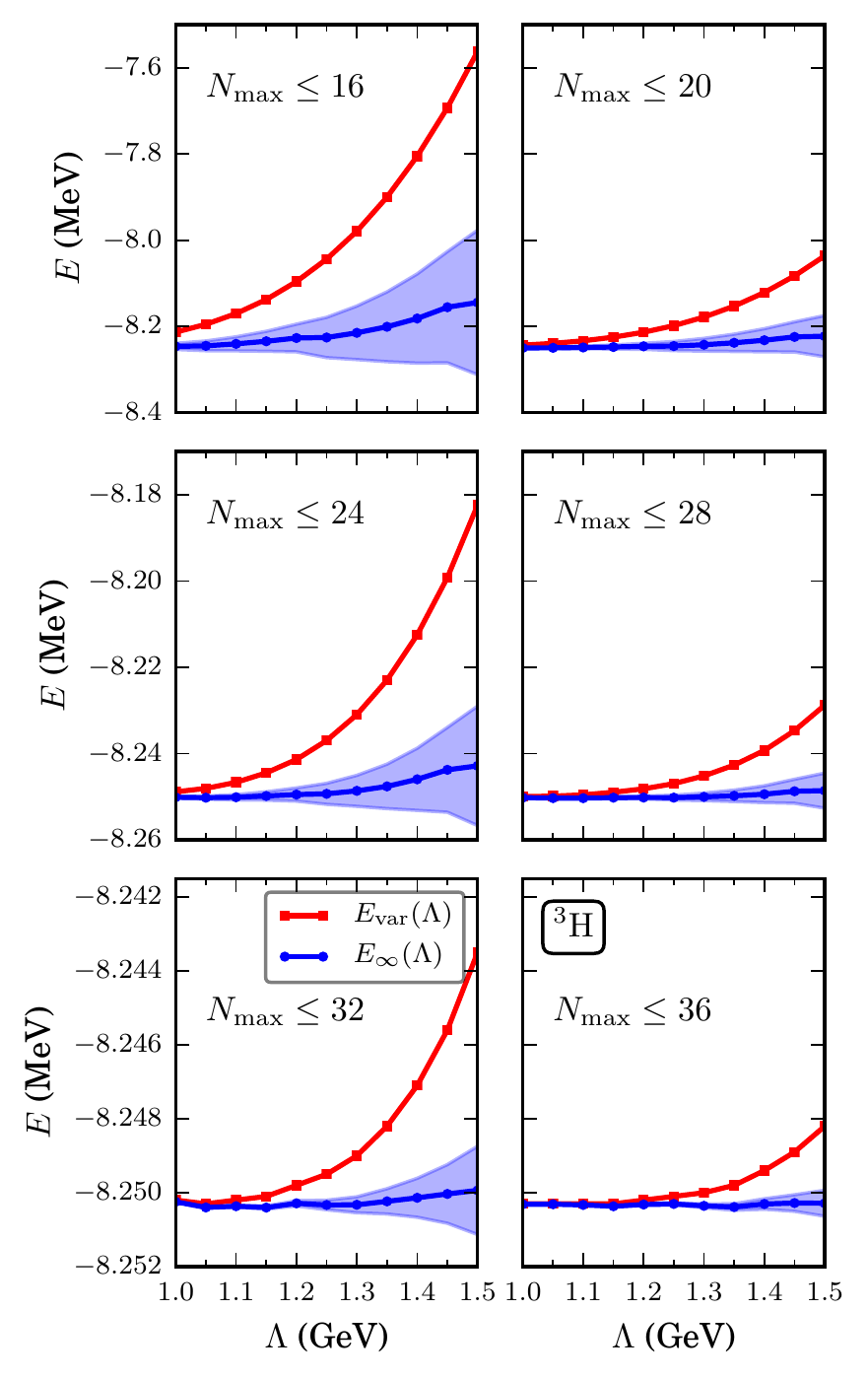}
  \caption{(Color online) Extrapolated energy $E_\infty(\Lambda)$
    (circles) for
    \nuc{3}{H} with the NNLO$_{\rm opt}$ $NN$ interaction. The different panels
    correspond to different NCSM model space truncations from
    $\max(N_\mathrm{max})=16$ to $\max(N_\mathrm{max})=36$. The 
    bands estimate uncertainties from subleading IR corrections. The
    squares denote the minimum energy computed with the NCSM as a
    function of $\Lambda$. }
  \label{fig:H3E0LO}
\end{figure}

Let us discuss first the extrapolation of energies.  The $\chi^2$ fits
of Eq.~(\ref{master}) to computational data employ the
uncertainty~(\ref{sigma}).  This uncertainty is a naive estimate of
subleading corrections to Eq.~(\ref{master}) and ensures that
numerical data is weighted correctly as a function of $L$.  The
results for \nuc{3}{H} energies are shown in Fig.~\ref{fig:H3E0LO}.
The squares show the variational minimum of the computed energy as a
function of the UV cutoff and for a given $\nm$. The extrapolated
results are shown as circles, with uncertainty
estimates given by Eq.~(\ref{sigma}), scaled with the extrapolation distance,
presented as a band.

We see that the extrapolated results are a significant improvement
over the NCSM results; with increasing $\nm{}$ they stabilize and are
constant over an increasing range of UV cutoffs. We note that the
uncertainties only estimate higher-order IR corrections due to the
first open decay channel. Missing UV corrections, or IR corrections
from other decay channels (with a separation momentum $k_\mathrm{sep}
>k_\infty$), are not included. For the triton, for instance, the
separation into three nucleons has a separation momentum
$k_\mathrm{sep}(t \to p+n+n) \approx 0.63$~fm$^{-1}$. This momentum is
not much larger than the separation momentum for the disintegration
$t\to d+n$.  We note that the displayed uncertainties increase with
increasing $\Lambda$, because at fixed $\nm{}$ the IR length $L$
decreases with increasing $\Lambda$.

\begin{figure}[htbp]
  \centering
  \includegraphics[width=\columnwidth]{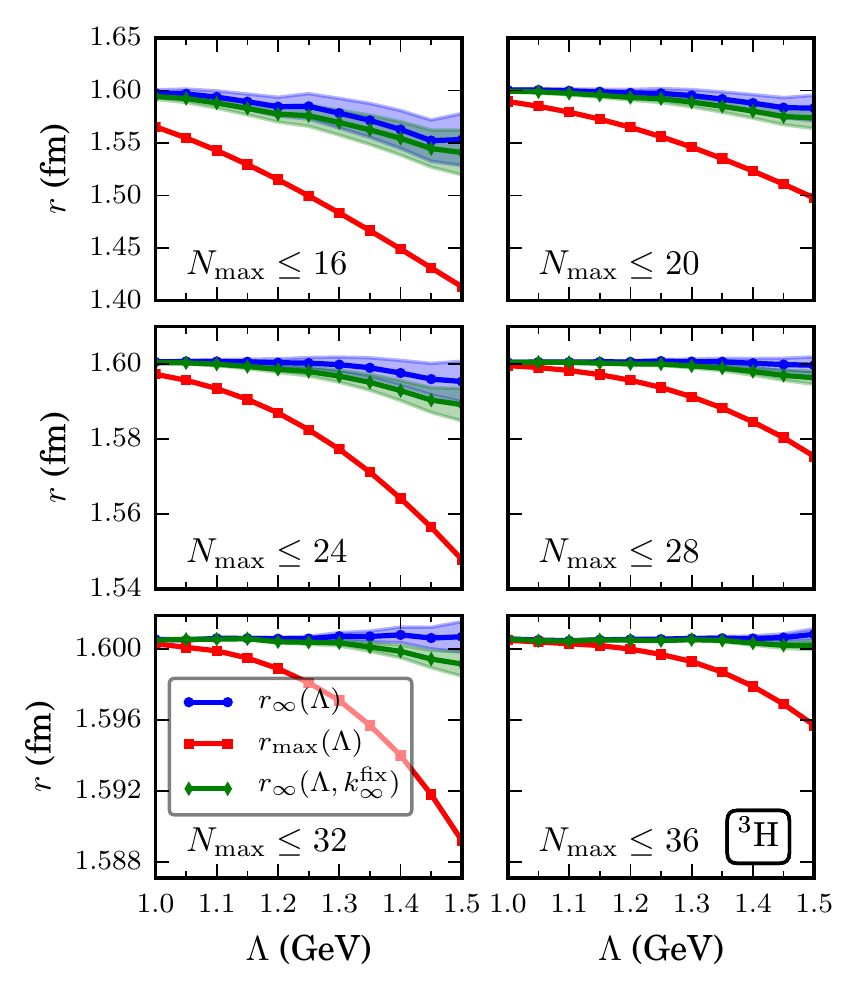}
  \caption{(Color online) Extrapolated ground-state (point-proton)
    radii $r_\infty(\Lambda)$
    (circles) for
    \nuc{3}{H} with the NNLO$_{\rm opt}$ $NN$ interaction. The different panels
    correspond to different NCSM model space truncations from
    $\max(N_\mathrm{max})=16$ to $\max(N_\mathrm{max})=36$. The 
    bands estimate uncertainties from subleading IR corrections. The
    squares denote the maximum radius computed with the NCSM as a
    function of $\Lambda$ and model space truncation. }
  \label{fig:H3RLO}
\end{figure}

The results for the extrapolated point-proton radius are presented in
Fig.~\ref{fig:H3RLO} as circles and compared to the values obtained
from the NCSM calculations.  Here, diamonds show extrapolation results
when $k_\infty$ is fixed from the energy extrapolation.  These resulted
in the reproduction of the exact ground-state energies in the interval
1000~MeV~$\lesssim \Lambda \lesssim$~1300~MeV.  In general, the
extrapolation that leaves $k_\infty$ as an adjustable parameter
(circles) yields more stable extrapolated radii, and the extrapolated
radius can be read off the plateau that develops as $N_{\rm max}$ is
increased.  In the $\chi^2$ fits of the radius, we use the uncertainty
(\ref{sigmar}) to account for subleading corrections. These
uncertainties, scaled with the extrapolation distance, are also shown
as bands in Fig.~\ref{fig:H3RLO}.

\begin{figure}[htbp]
  \centering
  \includegraphics[width=\columnwidth]{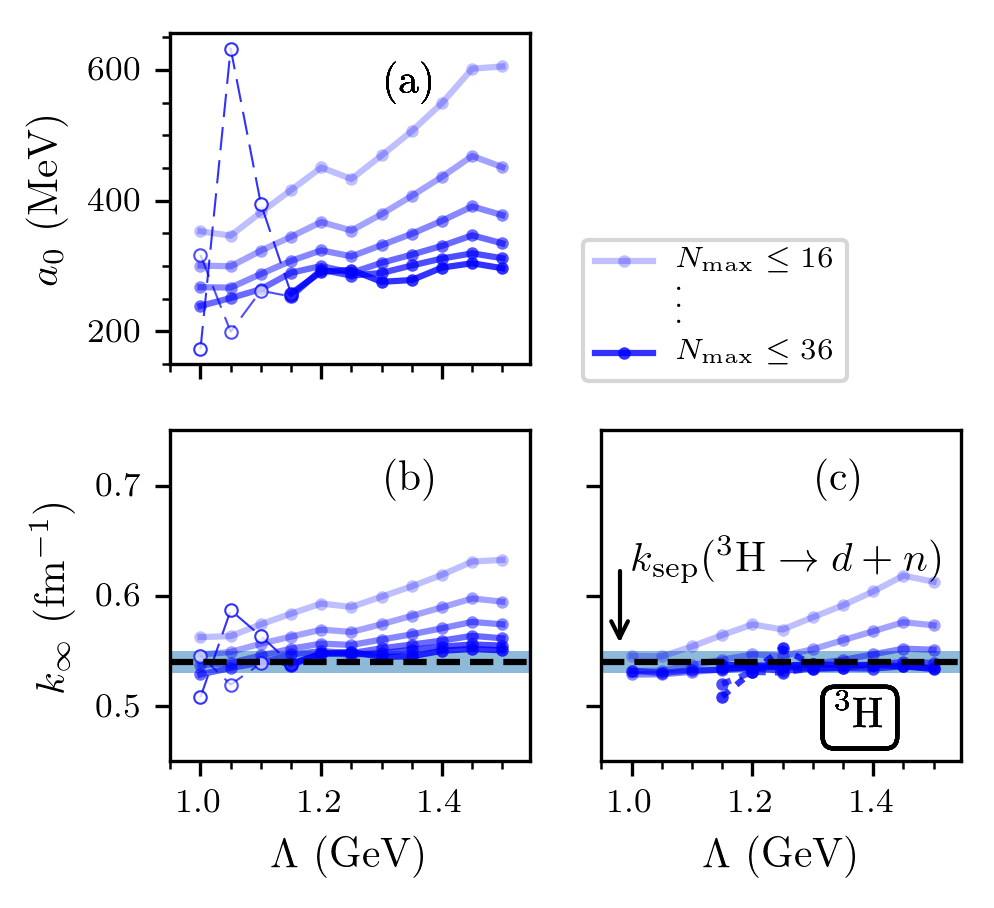}
  \caption{(Color online) Fit parameters $a_0$ (a),
    $k_\infty(\Lambda)$ for \nuc{3}{H} energy
    extrapolation (b), and $k_\infty(\Lambda)$ for radius extrapolation
    (c) for different NCSM model space truncations
    from $\max(N_\mathrm{max})=16$ to $\max(N_\mathrm{max})=36$.
    Open symbols denote results for which UV corrections are expected
    to be larger than IR ones, and the corresponding fits are unreliable.
    The lowest, theoretical separation
    momentum is given as a dashed line with an uncertainty band. }
  \label{fig:H3kLO}
\end{figure}

The values for $k_\infty$ resulting from the fit of Eq.~(\ref{master})
are shown in Fig.~\ref{fig:H3kLO}(b). We find that a
stable region is reached for large enough UV scales.  We note that
fits performed at a UV cutoff $\Lambda$ below the variational minimum
have UV corrections that are larger than the IR corrections. This is
reflected in a $\Lambda$ dependence of the fit parameters $a_0$ and
$k_\infty$. The values for $k_\infty$ obtained from the radius
extrapolation are shown in Fig.~\ref{fig:H3kLO}(c). In
large model spaces, the values obtained from the fit of energies and
radii agree with each other. We present the average value of these two
fit parameters, $k_\infty = 0.54(1)$ (obtained at the largest \nm\ and
$\Lambda$), as the recommended result in Table~\ref{tab:recLO}. This
numerical result validates our derivation in Sec.~\ref{deriv} since
the momentum scale extracted from the fits agrees very well with the
separation momentum $k_\mathrm{sep}(\nuc{3}{H})\approx 0.54$~fm$^{-1}$
obtained from the computed binding energies with the NNLO$_{\rm opt}$
interaction. \\

Values of $a_0$ from the fit to Eq.~(\ref{master}) are shown in
Fig.~\ref{fig:H3kLO}(a). For the largest \nm\ and $\Lambda$ we find
$a_0\approx 280$~MeV, which corresponds to the ANC
$\gamma_{\rm sep} \approx 3.5$~fm$^{-1/2}$ in the orthogonal Jacobi
coordinate $\rho_1$.  Using the results of Appendix~\ref{App_ANC}, the
ANC in the physical separation coordinate then becomes
$(2/3)^{1/4}\gamma_{\rm sep}\approx 3.2$~fm$^{-1/2}$.  We compare this
value with the experimental data of Refs.~\cite{kim1974,girard1979},
noting that experiments provide us with a dimensionless normalization
parameter. We use Eq.~(19) in the review~\cite{kim1974} to convert
this to an ANC of $2.1 - 3.4$~fm$^{-1/2}$, in agreement with our
theoretical value extracted from the fit.

These numerical results suggest that the relevant low-momentum scale
for a bound state in a many-body system indeed is set by the momentum
corresponding to the smallest separation energy. We note that this
conclusion is not limited to the oscillator basis, as similar results
were found for the lattice~\cite{koenig2018}. Of course, this is
consistent with view on the ANC, which governs astrophysical reaction
rates~\cite{mukhamedzhanov1990} at lowest energies.

We use the extrapolations at fixed $\Lambda=1200$~MeV to extract a
sequence of recommended values for the ground-state energy and the
point-proton radius for \nuc{3}{H} as a function of the model-space
truncation; see Fig.~\ref{fig:H3recLO}. For \nuc{3}{He} we find
results of similar quality; they are given in Appendix~\ref{sec:He3}.

\begin{figure}[htbp]
  \centering
  \includegraphics[width=\columnwidth]{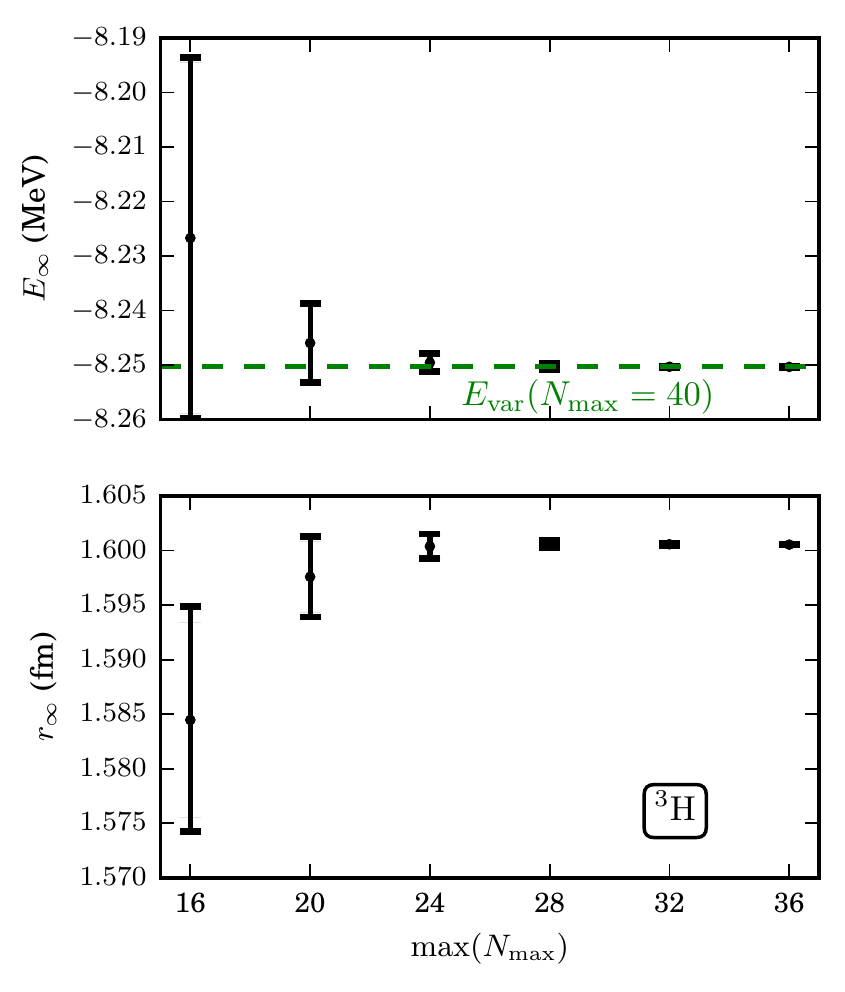}
  \caption{(Color online) Recommended results for the
    \nuc{3}{H} energy (upper panel) and radius
     (lower panel) for different NCSM model space
    truncations from $\max(N_\mathrm{max})=16$ to
    $\max(N_\mathrm{max})=36$.}
  \label{fig:H3recLO}
\end{figure}

\begin{table}[htb]
  \caption{Recommended results for the
    ground-state energy $E_\infty$ (in MeV) and point-proton radius $r_\infty$
     (in fm) for different nuclei. All results are obtained with the
    NNLO$_{\rm opt}$ $NN$ interaction. The variational minimum $E_{\rm
      var}^{\rm min}$ for each nucleus computed at the largest \nm\ reached in the NCSM
    calculations is also shown. Finally, the momentum scale,
    $k_\infty$ (extracted from the energy and radius fits) is compared with the lowest separation momentum,
    $k_\mathrm{sep}$, for this interaction from Eq.~\eqref{eq:sep}. 
    \label{tab:recLO}}
\begin{ruledtabular}
\begin{tabular}{c|d{3.4}d{1.3}d{4.2}cd{0.3}d{1.3}}
      & \multicolumn{1}{c}{$E_\infty$}
      & \multicolumn{1}{c}{$r_\infty$} 
      & \multicolumn{1}{r}{$E_{\rm var}^{\rm min}$}   
      & \multicolumn{1}{c}{$N_\mathrm{max}$}   
      & \multicolumn{1}{r}{$k_\infty$}
      & \multicolumn{1}{r}{$k_\mathrm{sep}$} \\ 
      \hline
      \nuc{3}{H}   & -8.250     & 1.60  & -8.250   & 40 
      & 0.54(1) & 0.54(1)\\
      \nuc{3}{He} & -7.502     & 1.793 & -7.502   & 40 
      & 0.51(2) & 0.51(1) \\
      \nuc{4}{He} & -27.592   & 1.434 & -27.592 & 20 
      & 0.84(5) & 0.97(3) \\
      \nuc{6}{Li}  & -30.59(3)  & 2.42(2) & -30.500 & 22 
      & 0.44(5) & 0.19(8) \\
      \nuc{6}{He} & -27.3(2)   & 1.84(2) & -26.976 & 16 
      & 0.47(3) & \multicolumn{1}{c}{--} \\   
      \nuc{8}{He} & -26.5(1.1) & 1.82(3) & -24.631 & 12 
      & 0.42(3) & \multicolumn{1}{c}{--} \\
      \hline
\end{tabular}
\end{ruledtabular}
\end{table}

Below, we will see that Eq.~(\ref{eq:kinf}) is also
semi-quantitatively fulfilled in $A=4,6$ and 16-body systems.

\subsection{\nuc{4}{He}
\label{subsec:A4}}

\begin{figure}[tbp]
  \centering
  \includegraphics[width=\columnwidth]{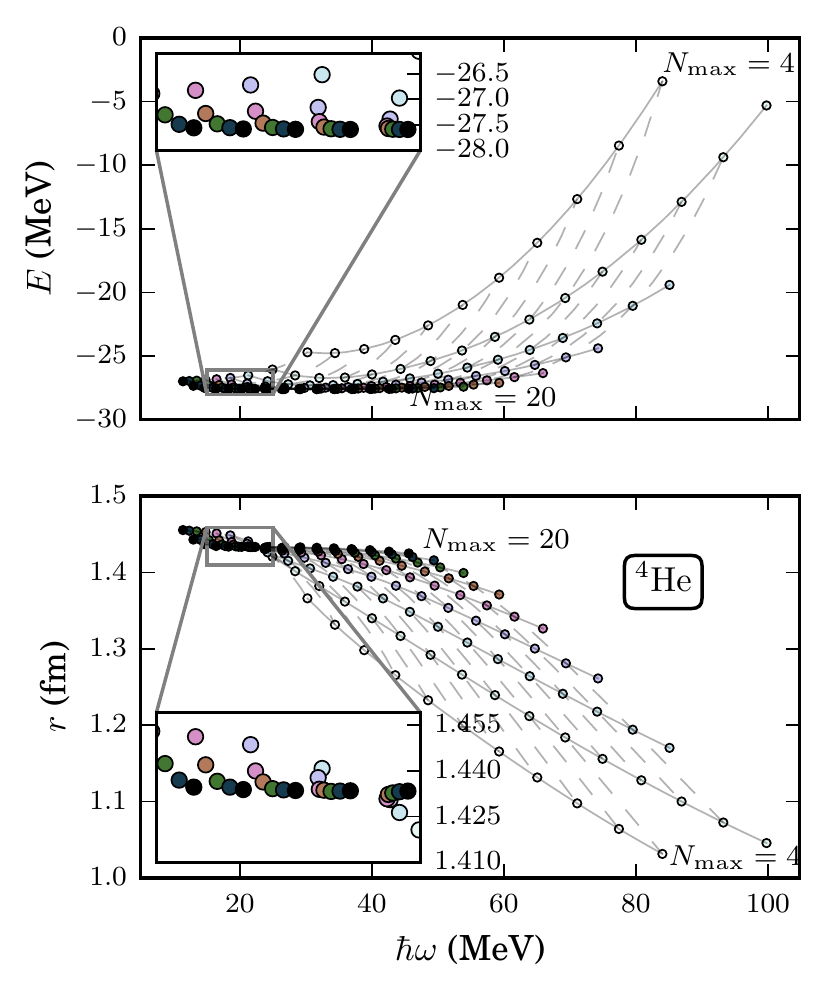}
  \caption{(Color online) Computed ground-state energies (upper panel)
    and point-proton radii (lower panel) for \nuc{4}{He} as
    a function of the oscillator spacing $\ho$ in model spaces
    of size $\nm$ as indicated. Solid
    lines connect data points with equal $\nm$. Dashed lines
    connect data points with equal UV cutoff $\Lambda$, starting at
    $\Lambda=750$~MeV to $\Lambda=1450$~MeV (from left
    to right in steps of 50~MeV). }
  \label{4Hedata}
\end{figure}

In this Section we present the IR extrapolations for the ground-state
energy and point-proton radius of \nuc{4}{He}.
The top panel of Fig.~\ref{4Hedata} shows the ground-state energies
for \nuc{4}{He} in model spaces with $\nm=4,6,\ldots,20$ as a
function of the oscillator spacing $\ho$.  Solid lines connect data
points with equal $\nm$. Dashed lines connect data points with equal
$\Lambda$, starting at $\Lambda=750$~MeV to $\Lambda=1500$~MeV (from
left to right in steps of 50~MeV). In what follows, we will perform IR
extrapolations with Eq.~(\ref{master}) based on data points computed
in model spaces with equal UV cutoff $\Lambda$.

The lower panel of Fig.~\ref{4Hedata} shows the computed ground-state
point-proton radius for \nuc{4}{He} as a function of the oscillator
spacing for model spaces of size $\nm$ as indicated.  Solid lines
again connect radii at fixed $\nm$ while dashed lines connect data at
fixed UV cutoff $\Lambda$. The results become almost independent of
$\nm$ around $\hbar\omega\approx 23$~MeV, and it makes sense to
identify this value as the theoretical radius in an infinite space;
see, e.g., Refs.~\cite{bogner2008,caprio2012,shin2017}. Below we will
see that the radius extrapolations yield plateaus that allow one to
read off the radius with more confidence also when no full convergence can be achieved.

The trend of the radius curves can be understood as follows. With
increasing oscillator spacings, the computed radius decreases because
the IR length of the model space also decreases. In this regime, UV
corrections to the bound-state become increasingly smaller. For
decreasing values of the oscillator spacing, the UV cutoff $\Lambda$
decreases, and the computed binding energy decreases, thus leading to
a more weakly bound system and an ever increasing radius. In this
regime, IR corrections to the radius become increasingly smaller as
the oscillator spacing is further decreased.

\begin{figure}[tbp]
  \centering
  \includegraphics[width=\columnwidth]{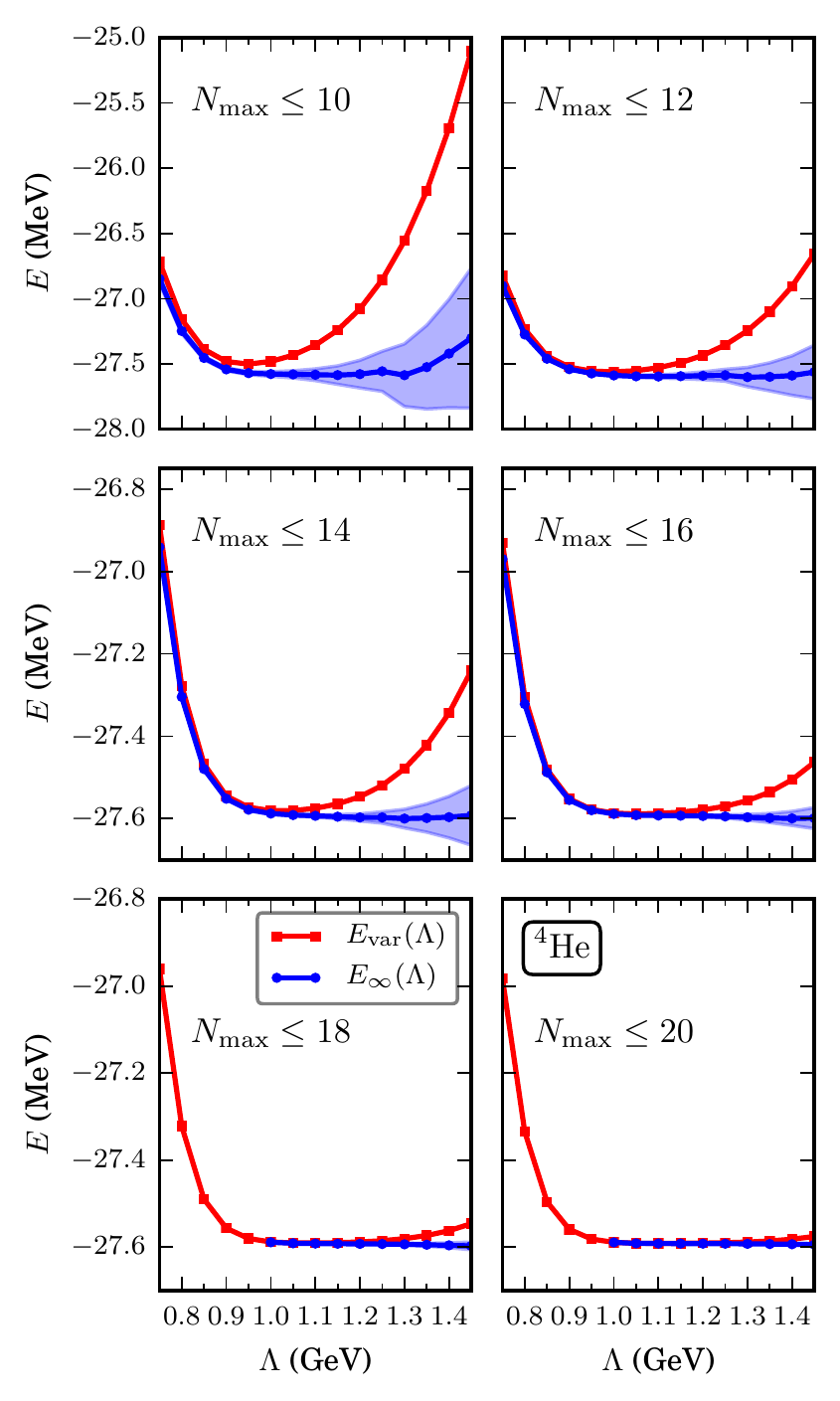}
  \caption{(Color online) Extrapolated energy $E_\infty(\Lambda)$
    (circles) for
    \nuc{4}{He}. The different panels
    correspond to different NCSM model space truncations from
    $\max(N_\mathrm{max})=10$ to $\max(N_\mathrm{max})=20$. 
    See caption of Fig. \ref{fig:H3E0LO} for further details.
  }
  \label{He4E0LO}
\end{figure}

We perform a $\chi^2$ fit to the ground-state energies $E(L,\Lambda)$
based on the extrapolation formula~(\ref{master}) and use the
theoretical uncertainties~(\ref{sigma}) in the fit.  We recall that
this uncertainty only accounts for some of the missing IR
corrections. Missing UV corrections are not addressed and one should
therefore not expect a proper error estimate for small values of
$\Lambda$. The fit results for the parameters $E_\infty(\Lambda)$
are shown as circles in
Fig.~\ref{He4E0LO} for various values of $\nm$. We see that the
extrapolation energies are an improvement compared to the variational
minima (shown as squares).

\begin{figure}[tbp]
  \centering
  \includegraphics[width=\columnwidth]{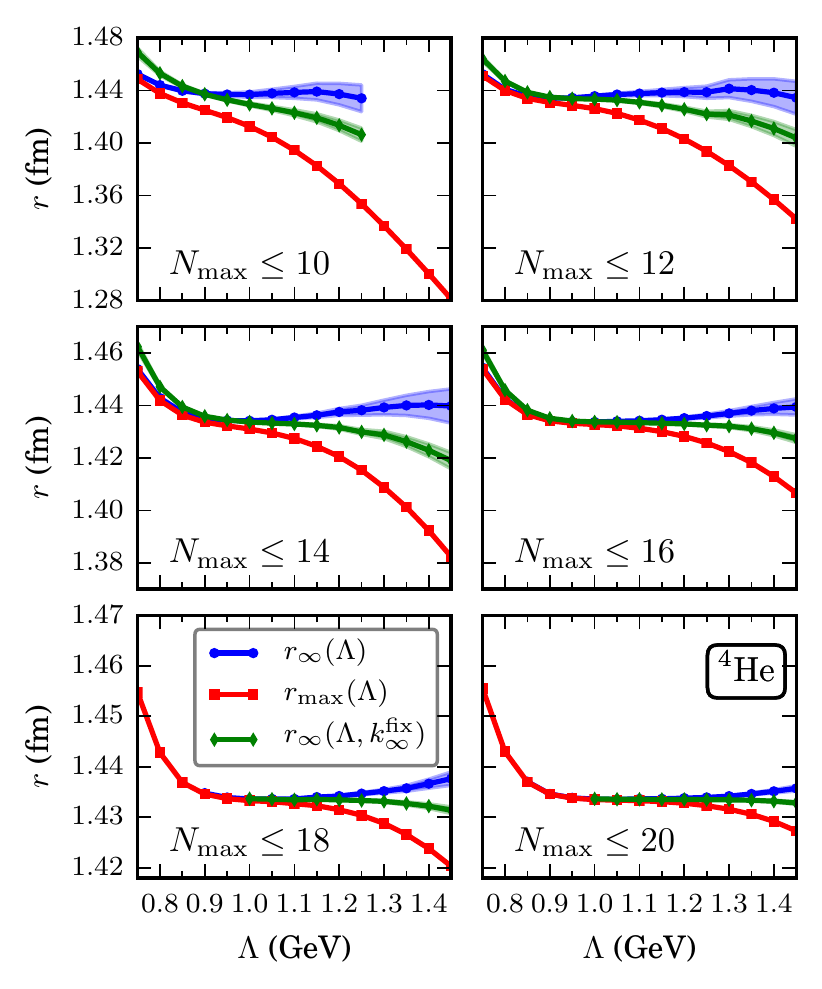}
  \caption{(Color online) Extrapolated ground-state (point-proton)
    radii $r_\infty(\Lambda)$
    (circles) for
    \nuc{4}{He}. The different panels
    correspond to different NCSM model space truncations from
    $\max(N_\mathrm{max})=10$ to $\max(N_\mathrm{max})=20$.
    See caption of Fig. \ref{fig:H3RLO} for further details.
  }
  \label{fig:He4RLO}
\end{figure}

For the ground-state radius we perform $\chi^2$ fits of
Eq.~(\ref{radii}) to our computed results, using the
uncertainty~(\ref{sigmar}) to account for subleading corrections. The
results for $r_\infty$ and the corresponding uncertainty estimates are
shown as circles and bands in Fig.~\ref{fig:He4RLO}; here we used
$k_\infty$ as a fit parameter. In contrast, one might also employ for
$k_\infty$ the same values as found in the energy extrapolation.
Employing the latter in the fit of the radii [i.e., making only
  $r_\infty$ and $\alpha$ adjustable parameters in Eq.~(\ref{radii})]
yields extrapolated results that are shown as diamonds in
Fig.~\ref{fig:He4RLO}, with a green uncertainty band. In very large
spaces, both extrapolation results approach each other. In smaller
spaces, extrapolated radii exhibit a weaker $\Lambda$ dependence if
$k_\infty$ is an adjustable parameter. The extrapolation results can
be compared to the computed NCSM results (squares).

\begin{figure}[tbp]
  \centering
  \includegraphics[width=\columnwidth]{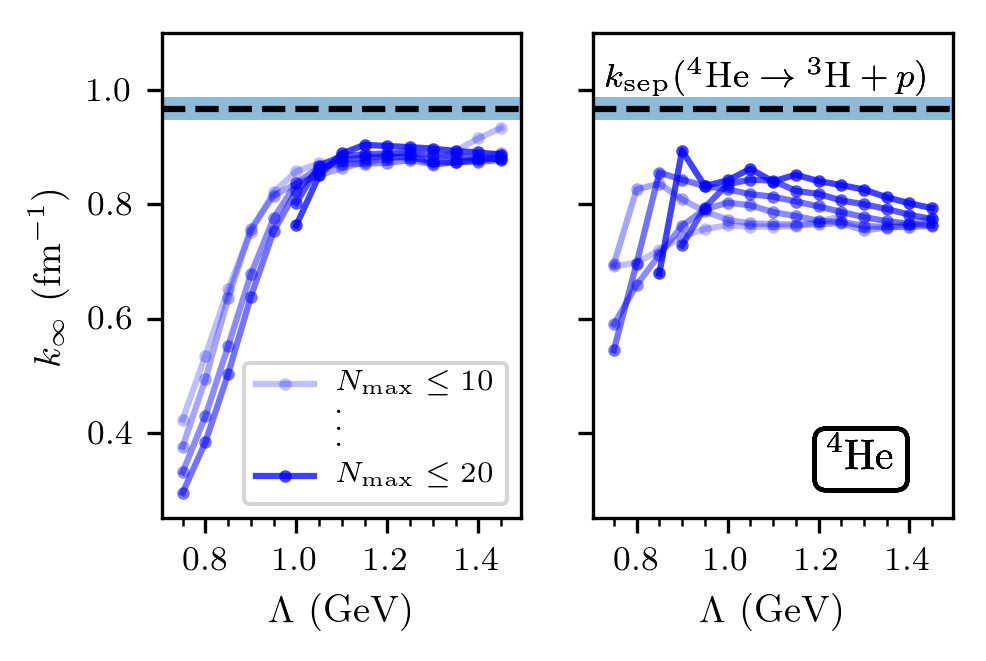}
  \caption{(Color online) Fit parameter $k_\infty(\Lambda)$ for
    \nuc{4}{He} energy extrapolation (left panel) and radius
    extrapolation (right panel) for different NCSM model space
    truncations from $\max(N_\mathrm{max})=10$ to
    $\max(N_\mathrm{max})=20$. The lowest, theoretical separation
    momentum is given as a dashed line with an uncertainty band.}
  \label{He4kLO}
\end{figure}

The results for the fit parameter $k_\infty$ from the energy and
radius extrapolations are shown in the top and bottom panels of
Fig.~\ref{He4kLO}, respectively. For the largest model spaces and
cutoffs around 1~GeV they are consistent with (but not identical to)
each other. The $\nm$ dependence of $k_\infty$ is smallest for the
energy extrapolation, and we focus on them.  We note that $k_\infty$
depends weakly on $\Lambda$ as this quantity increases beyond
$\Lambda\gtrsim 1000$~MeV. This is consistent with our expectations
because these results are increasingly well UV converged.  We find
$k_\infty\approx 0.87\pm 0.03$~fm$^{-1}$ from the energy extrapolation
and compute a corresponding separation energy
$S = {k_\infty^2\over 2 m} \approx 15.8\pm 1.1$~MeV.
This value is somewhat smaller than the theoretical values for the
proton and neutron separation energies $S_p\approx 19.3$~MeV and
$S_n\approx 20.1$~MeV, respectively.  The corresponding separation
momenta are $k_\mathrm{sep} \approx 0.96$ and 0.98~fm$^{-1}$, see
Table~\ref{tab:recLO}, about 10\% larger than from our energy
extrapolation.

We note that the two-nucleon separation energies are significantly larger
than the nucleon separation energy of about 20~MeV. Thus, they should
yield only smaller and negligible corrections. Furthermore, the
\nuc{4}{He} nucleus is essentially an $s$-wave state, and corrections
to the energy extrapolation~(\ref{IR}) due to other partial waves are
also expected to be small.

\begin{figure}[tbp]
  \centering
  \includegraphics[width=\columnwidth]{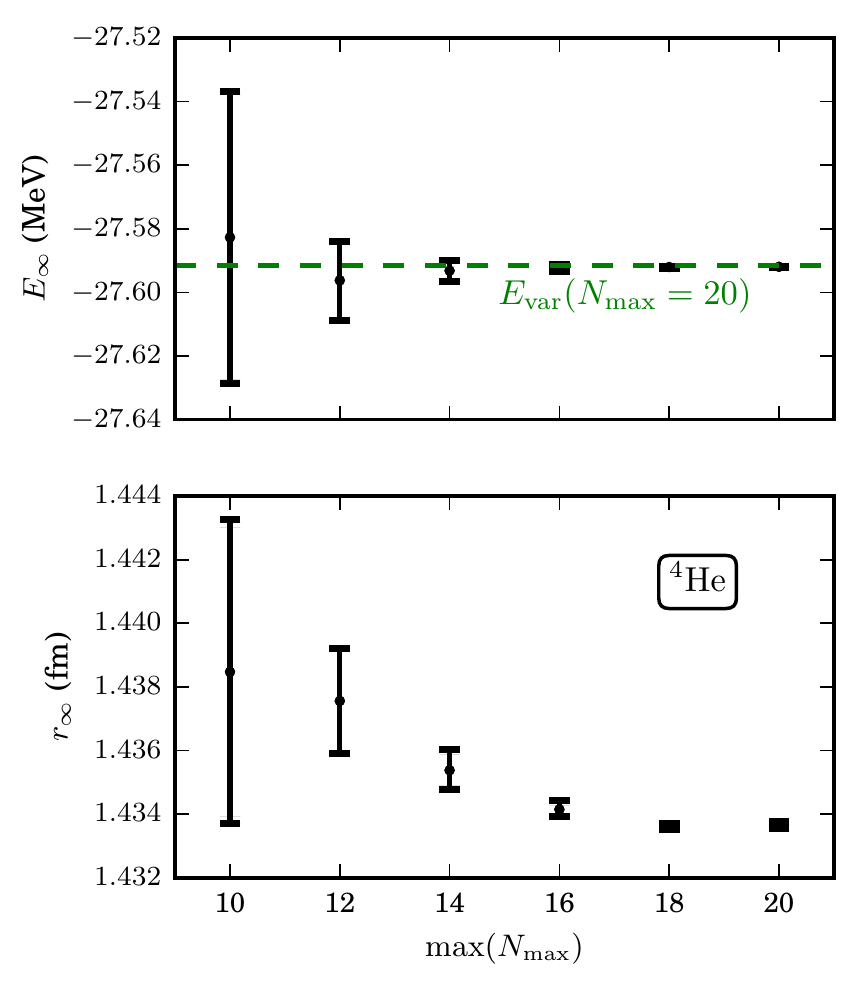}
  \caption{(Color online) Recommended results for the
    \nuc{4}{He} energy (upper panel) and radius
     (lower panel) for different NCSM model space
    truncations from $\max(N_\mathrm{max})=10$ to
    $\max(N_\mathrm{max})=20$.}
  \label{He4recLO}
\end{figure}

We collect recommended values for ground-state energy and the charge
radius of \nuc{4}{He} in the panels of Fig.~\ref{He4recLO}. The
extrapolated values and corresponding uncertainties are taken at
$\Lambda=1100$~MeV.

\section{$p$-shell nuclei: Low-momentum scales of many-body systems
\label{sec:pshell}}
\subsection{\nuc{6}{Li}
\label{subsec:A6}}

\nuc{6}{Li} is a weakly bound nucleus due to the proximity of the
$\nuc{6}{Li}\to\nuc{4}{He} + d$ breakup channel at only 1.5~MeV
excitation energy. The experimental charge radii $r_c(\nuc{A}{Li}) =
2.5432(262)$, 2.4173(28), and 2.327(30)~fm for $A=6$, 7, and 8,
respectively~\cite{sanchez2006}, confirm that \nuc{6}{Li} can be
viewed as a deuteron-halo nucleus. This makes {\it ab initio}
computation of this nucleus somewhat
challenging~\cite{pudliner1997,dytrych2015,shin2017}, and IR
extrapolations can be useful.

\begin{figure}[tbp]
  \centering
  \includegraphics[width=\columnwidth]{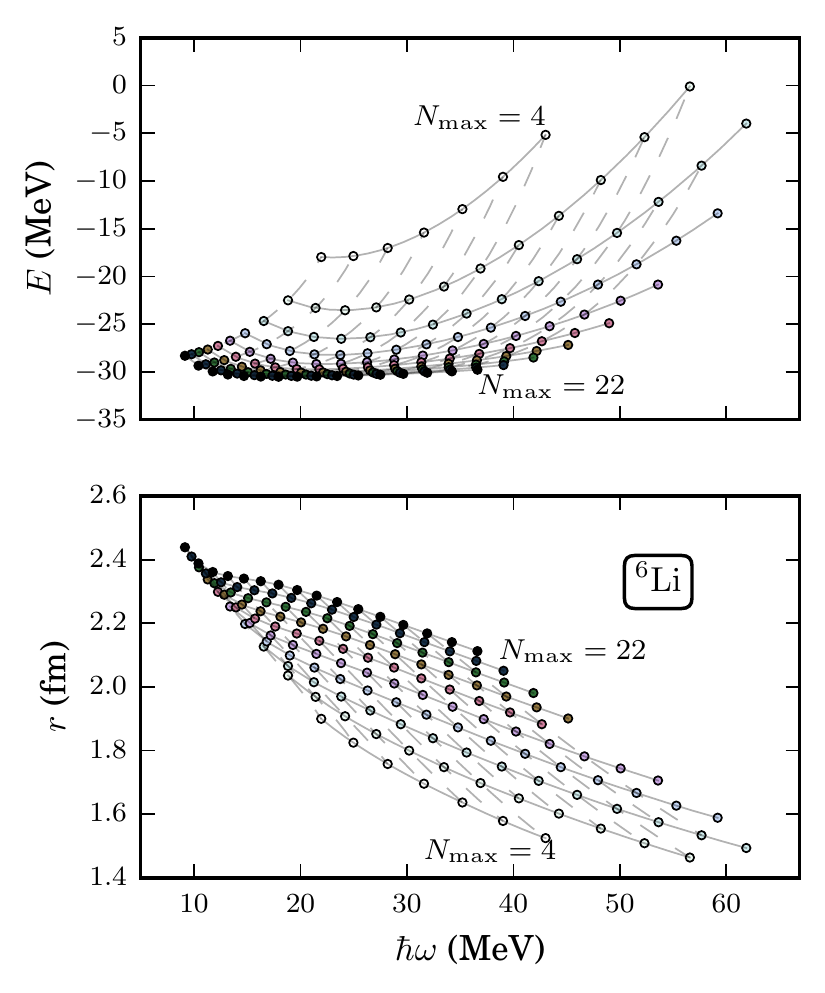}
  \caption{(Color online) Computed ground-state energies (upper panel)
    and point-proton radii (lower panel) for \nuc{6}{Li} as
    a function of the oscillator spacing $\ho$ in model spaces
    of size $\nm$ as indicated. Solid
    lines connect data points with equal $\nm$. Dashed lines
    connect data points with equal UV cutoff $\Lambda$, starting at
    $\Lambda=750$~MeV to $\Lambda=1400$~MeV (from left
    to right in steps of 50~MeV). }
  \label{6Lidata}
  \end{figure}

At fixed UV cutoff $\Lambda$ and $N$ we compute the corresponding
oscillator spacing $\ho(N,\Lambda)$, using the tables presented in
Ref.~\cite{wendt2015} for the nucleus \nuc{6}{Li}. We choose model spaces
with $N\le \nm$ and compute the ground-state energy and its
point-proton radius.
The upper panel of Fig.~\ref{6Lidata} shows the ground-state energies for \nuc{6}{Li} in
model spaces with $\nm=4,6,8\ldots,22$ as a function of the oscillator
spacing $\ho$.  

For the radii in the lower panel, many lines merge around $r\approx
2.3$~fm at $\ho\approx 12$~MeV, and one might be tempted to identify
this almost-$\nm$-independent value as the theoretical radius in an
infinite space. As we will see below, however, IR extrapolations yield
a larger radius than this merging point might suggest.

\begin{figure}[htbp]
  \centering
  \includegraphics[width=\columnwidth]{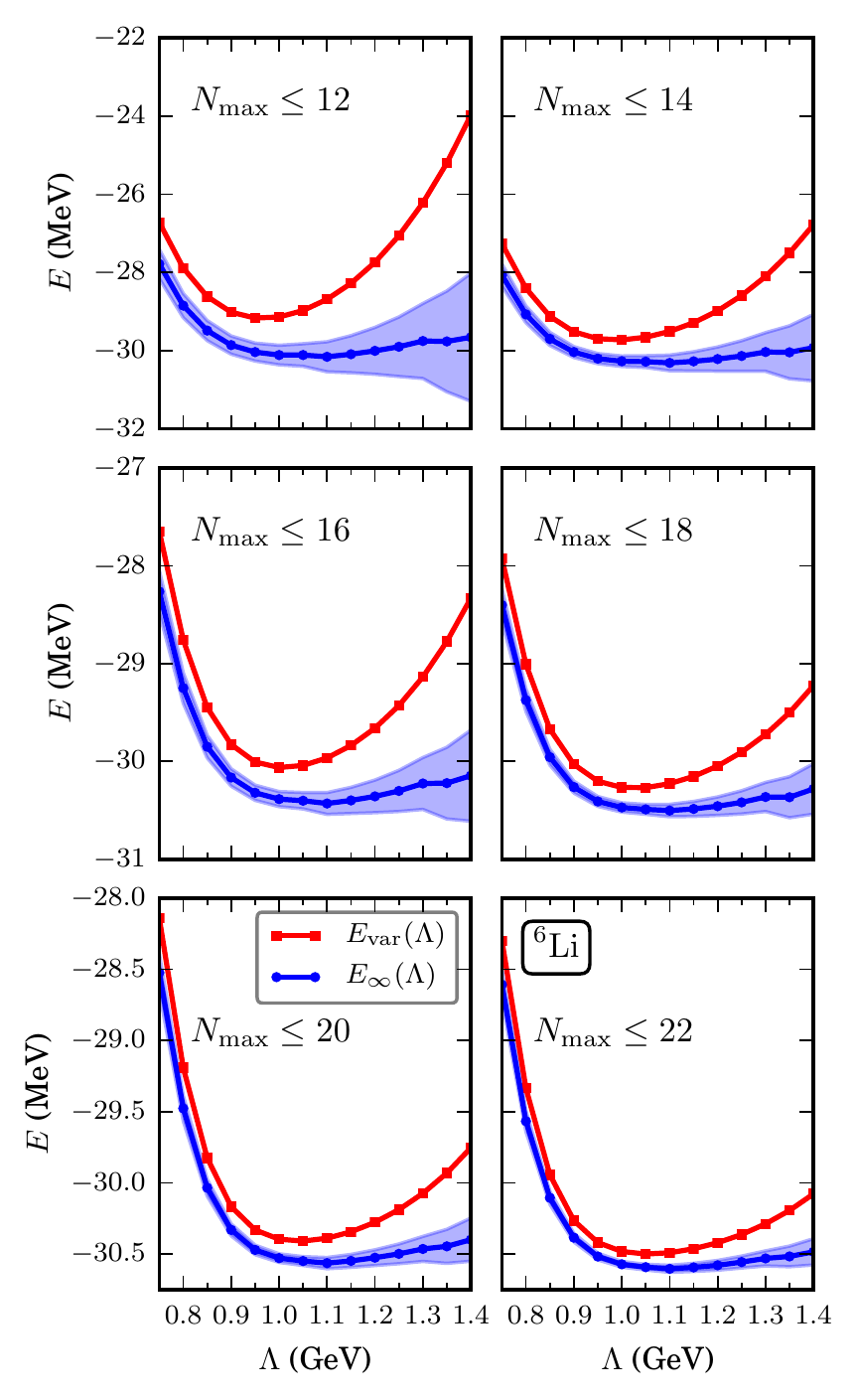}
  \caption{(Color online) Extrapolated energy $E_\infty(\Lambda)$
    (circles) for
    \nuc{6}{Li}. The different panels
    correspond to different NCSM model space truncations from
    $\max(N_\mathrm{max})=12$ to $\max(N_\mathrm{max})=22$. 
    See caption of Fig. \ref{fig:H3E0LO} for further details.
  }
  \label{Li6E0LO}
\end{figure}

We perform a $\chi^2$ fit to the resulting energies based on the
extrapolation formula~(\ref{master}) and use the theoretical
uncertainties~(\ref{sigma}).  The fit results for the energy
$E_\infty(\Lambda)$ are shown in Fig.~\ref{Li6E0LO}.  We see that for
a range of $\Lambda$ around 1100~MeV, the IR extrapolation
significantly improves over the variational minimum.  For smaller
values of $\Lambda\lesssim1000$~MeV (but still above the UV cutoff of
NNLO$_{\rm opt}$ $\Lambda_\chi=500$~MeV), the lack of UV convergence
yields energies that increase with decreasing values of $\Lambda$.
For larger values of $\Lambda\gtrsim 1200$~MeV the extrapolated
energies increase as $\Lambda$ is increased. This can be understood as
follows: At large values of $\Lambda$, UV convergence is presumably
achieved. However, as is reflected by the increasing theoretical
uncertainties, the maximum $L$ values achievable in spaces with $N\le
\nm$ decrease with increasing $\Lambda$, and we are not any more in
the regime where the asymptotic formula~(\ref{master}) is valid. We
recall that at smaller values of $L$ subleading IR corrections to
Eq.~(\ref{master}) must become relevant, leading to a more complicated
non-exponential (i.e.\ slower than exponential) IR
convergence. Fitting such data points with an exponential yields
higher values for $E_\infty$. These results at high values of
$\Lambda$ are consistent with Refs.~\cite{furnstahl2014,wendt2015}
where IR extrapolations for fully UV-converged ground-state energies
are close to the variational minimum, thus questioning the utility of
such extrapolations.

\begin{figure}[htbp]
  \centering
  \includegraphics[width=\columnwidth]{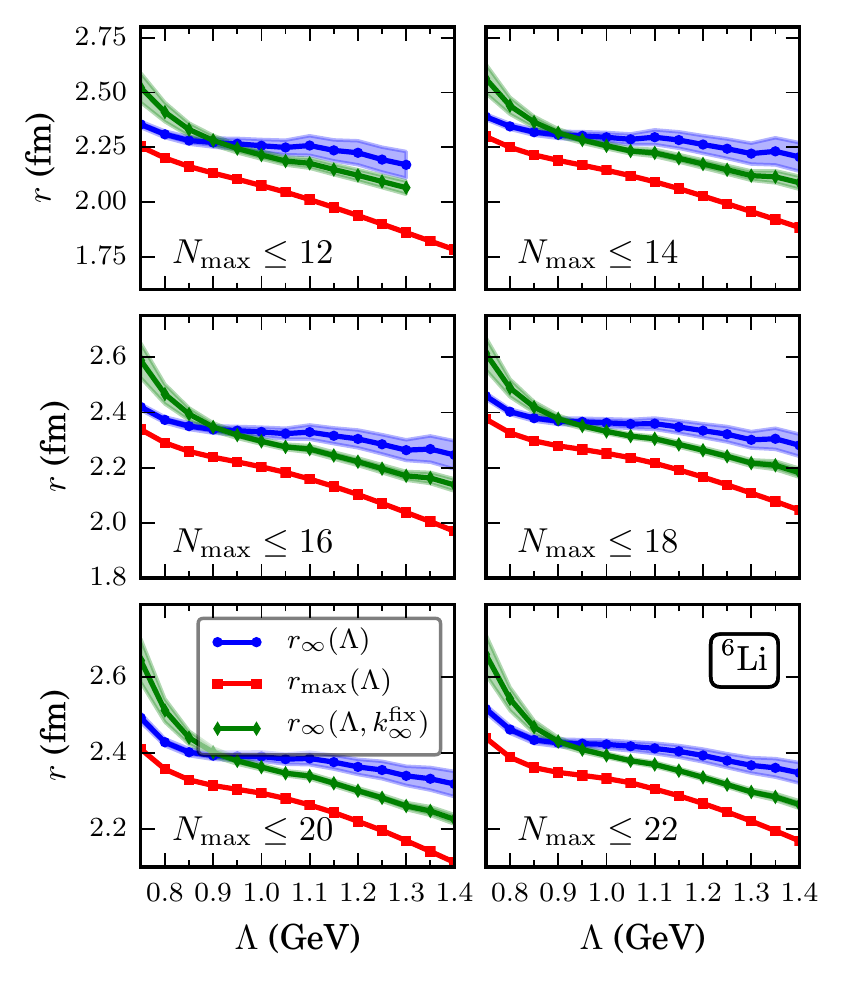}
  \caption{(Color online) Extrapolated ground-state (point-proton)
    radii $r_\infty(\Lambda)$
    (circles) for
    \nuc{6}{Li}. The different panels
    correspond to different NCSM model space truncations from
    $\max(N_\mathrm{max})=10$ to $\max(N_\mathrm{max})=20$. 
    See caption of Fig. \ref{fig:H3RLO} for further details.
  \label{fig:Li6RLO}}
\end{figure}

Our results for the point-proton radius are shown in
Fig.~\ref{fig:Li6RLO} for increasing values of $\nm$.  The raw results
$r(\nm,\Lambda)$ are shown as red squares. Blue circles show
extrapolated results where $k_\infty$ is treated as a fit parameter;
here, $\Lambda$-independent plateaus develop as $\nm$ is increased.
Green diamonds show extrapolation results where $k_\infty$ is taken
from the energy extrapolation; while the results display a weaker
$\Lambda$ dependence than the raw data, no plateaus are formed in this
case.  Our extrapolated radius result, $r_\infty= 2.417 \pm 0.02$~fm,
yields a charge radius $r_c=2.55(2)$~fm in agreement with
data~\cite{sanchez2006}, and the theoretical uncertainty also reflects
that our radius is not yet fully converged. Here we used the well-known
formula $r_c^2=r_\infty^2 +\langle r_p^2\rangle +(N/Z)\langle
r_n^2\rangle + 0.033$~fm$^2$, where $\langle
r_p^2\rangle=0.769$~fm$^2$ and $\langle r_n^2\rangle=-0.116$~fm$^2$
are the mean squared charge radii of the proton and neutron,
respectively, and the last correction is the Darwin-Foldy term.

\begin{figure}[tbp]
  \centering
  \includegraphics[width=\columnwidth]{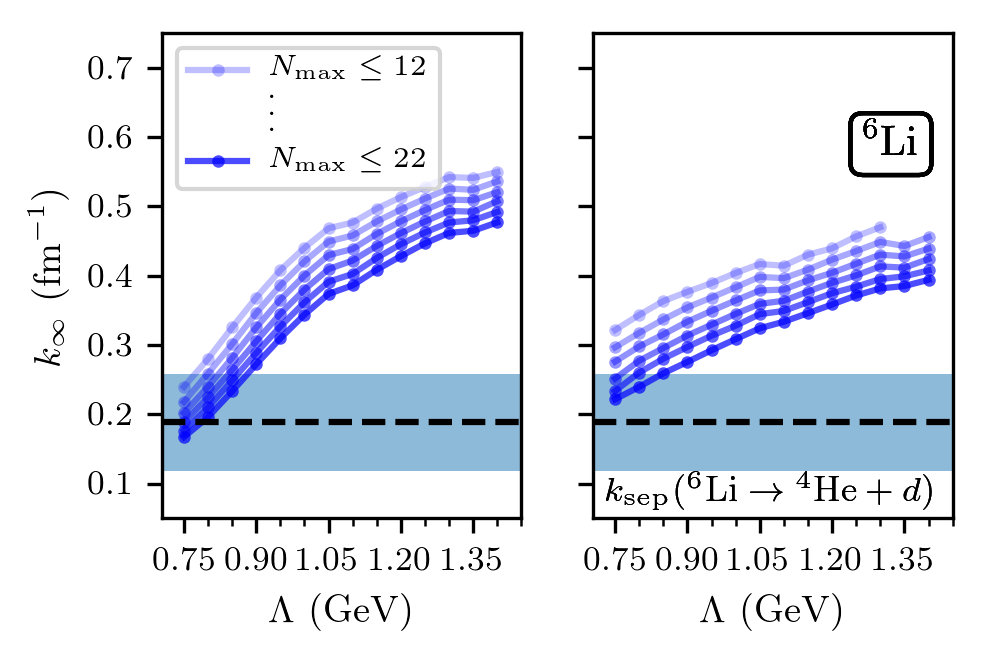}
  \caption{(Color online) Fit parameter $k_\infty(\Lambda)$ for
    \nuc{6}{Li} energy extrapolation (left panel) and radius
    extrapolation (right panel) for different NCSM model space
    truncations from $\max(N_\mathrm{max})=10$ to
    $\max(N_\mathrm{max})=20$. The lowest, theoretical separation
    momentum is given as a dashed line with an uncertainty band.}
  \label{Li6kLO}
\end{figure}

The momenta $k_\infty$ resulting from the energy and radius
extrapolation are shown in the left and right panel of
Fig.~\ref{Li6kLO}, respectively.  These momentum parameters only start
to stabilize for the largest values of $\Lambda$, perhaps suggesting
that UV convergence is about to be reached. We note that $k_\infty$
still decreases by about 0.02~fm$^{-1}$ as $\nm$ is increased from 14
to 16 (and from 16 to 18). The values from the energy and radius
extrapolation both lack IR convergence and differ by about 20\%.

For \nuc{6}{Li}, the deuteron separation energy (or the alpha-particle
separation energy) is $S_\alpha\approx 1.6$~MeV experimentally and
about 0.8~MeV when computed from the binding energy difference between
\nuc{6}{Li} on the one hand, and \nuc{4}{He} and \nuc{2}{H} on the
other hand.  This small separation energy makes the computation of the
binding energy and radius of this nucleus a challenge in {\it ab
  initio} calculations. Single-nucleon separation energies are
significantly larger, and corresponding IR corrections are thus much
smaller. At $\Lambda_{\rm UV}=1500$~MeV in the largest model spaces,
we have $k_\infty L\approx 4$, barely in the regime $k_\infty L\gg 1$
that is required for IR extrapolations.  As the \nuc{6}{Li} nucleus
has significant $s$-wave and $p$-wave contributions, the corresponding
$(k_\infty L)^{-1}$ corrections to the IR extrapolation
formula~(\ref{IR}) could still be sizable.  This is possibly also
reflected in the slow convergence of $k_\infty$. Taking a value of
$k_\infty\approx 0.44 \pm 0.05$~fm$^{-1}$ from the
extrapolations we find for the separation energy
\be
S_\alpha =  {\hbar^2 k_\infty^2\over 2 m} \approx 3.9\pm 1~\mbox{MeV.}
\ee
This is still larger than expected from binding-energy differences. We
note that our value for $k_\infty$ is consistent with the value
$k_\infty\approx 0.49$~fm$^{-1}$ reported in Ref.~\cite{wendt2015}.

\begin{figure}[tbp]
  \centering
  \includegraphics[width=\columnwidth]{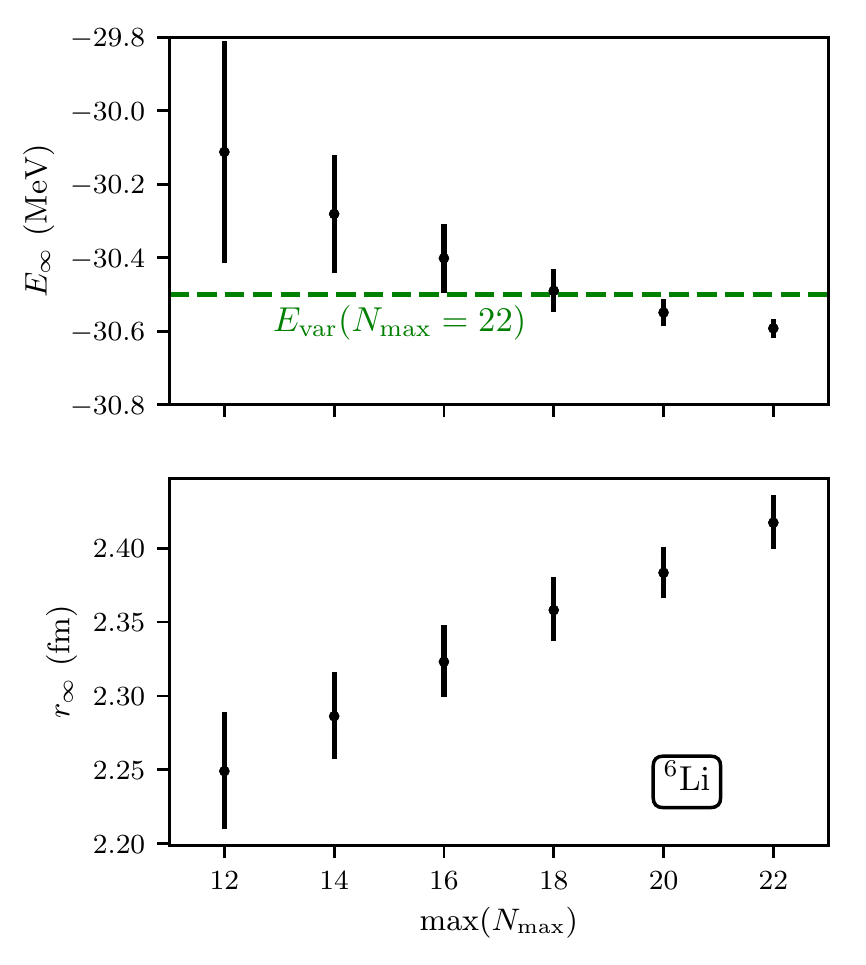}
  \caption{(Color online) Recommended results for the
    \nuc{6}{Li} energy (upper panel) and radius
     (lower panel) for different NCSM model space
    truncations from $\max(N_\mathrm{max})=12$ to
    $\max(N_\mathrm{max})=22$.}
  \label{Li6recLO}
\end{figure}

Recommended values for the ground-state energy of \nuc{6}{Li} are shown in
the top panel of Fig.~\ref{Li6recLO}. 

\subsection{\nuc{6,8}{He}
  \label{subsec:He}}
\begin{figure}[htbp]
  \centering
  \includegraphics[width=\columnwidth]{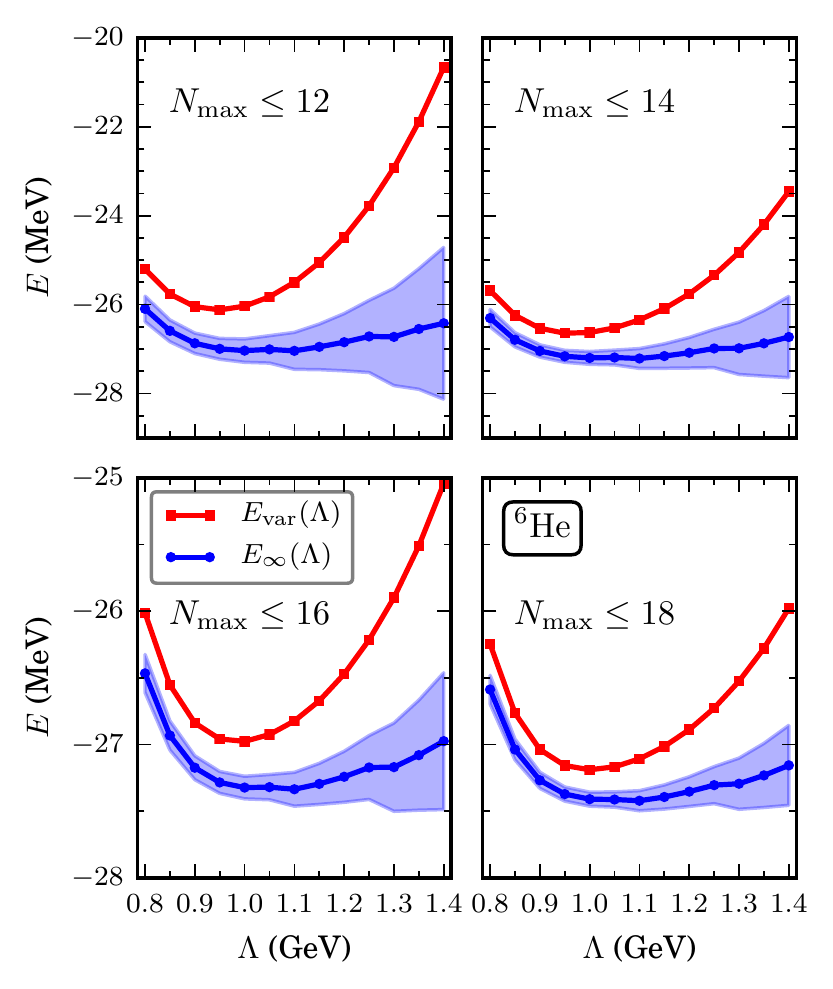}
  \caption{(Color online) Extrapolated ground-state energy
    $E_\infty(\Lambda)$ for
    \nuc{6}{He}. The different panels
    correspond to different NCSM model space truncations. See caption of Fig. \ref{fig:H3E0LO} for further details.
  \label{fig:He6ELO}}
\end{figure}

We also computed the nuclei \nuc{6,8}{He}. In finite model spaces, the
binding energies of these nuclei are smaller than the binding energy
of \nuc{4}{He}. Thus, they are unbound with respect to emission of the
alpha-particle. While this is a shortcoming of the employed NNLO$_{\rm opt}$
interaction, it is still interesting to study these cases in more
detail. In an infinite space, the \nuc{6,8}{He} systems are thus a \nuc{4}{He}
nucleus and free neutrons, and the expectation is that the
ground-state energy is that of the \nuc{4}{He} nucleus (as kinetic energies
of the neutrons can be arbitrarily small and the $s$-wave scattering
of neutrons among each other and off the \nuc{4}{He} nucleus yields
arbitrary small contributions). Thus, the wave functions of the
\nuc{6,8}{He} systems would not fall off exponentially, and the
extrapolation formulas could not be applied. However, when applying
the extrapolation formulas to these systems, we still got meaningful
results, i.e.\ the energy seems to converge exponentially with
increasing size $L$ of the model space. This unexpected result is
shown in Fig.~\ref{fig:He6ELO} for \nuc{6}{He}. 

How can this be understood? The model spaces we employ have a maximum
extent (i.e.\ a corresponding hard-wall radius) of about $L\approx
10$~fm in position space. The \nuc{6,8}{He} nuclei have nucleons with
angular momentum $l=1$ in a simple shell-model picture, and grand
angular momentum $K=2$ in hyperspherical coordinates. The
corresponding angular momentum barrier is $ \hbar^2 (K+3/2)(K+5/2) /
(2m L^2)\approx 3$~MeV high even at the hard-wall radius. (For \nuc{4}{He}
we have $K=0$ and the barrier is less than an MeV at the boundary.)
Thus, the binding of the \nuc{6}{He} nucleus is a transient behavior that
appears in model spaces that are sufficiently large to exhibit a
convergence of results but still too small to reflect the asymptotic
true nature of this six-nucleon system.

\begin{figure}[tbp]
  \centering
  \includegraphics[width=\columnwidth]{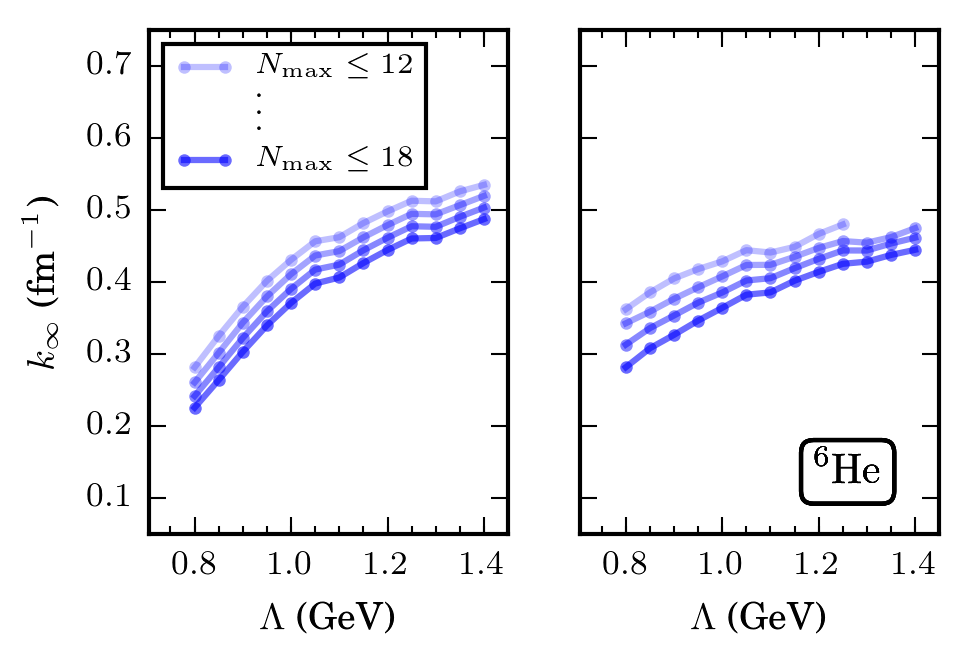}
  \caption{(Color online) Fit parameter $k_\infty(\Lambda)$ for
    \nuc{6}{He} energy extrapolation (left panel) and radius
    extrapolation (right panel) for different NCSM model space
    truncations from $\max(N_\mathrm{max})=12$ to
    $\max(N_\mathrm{max})=18$.}
  \label{He6kLO}
\end{figure}

The top panel of Fig.~\ref{He6kLO} shows the corresponding $k_\infty$
values obtained from the extrapolation of the ground state
energy. These values are still not converged.

\begin{figure}[htbp]
  \centering
  \includegraphics[width=\columnwidth]{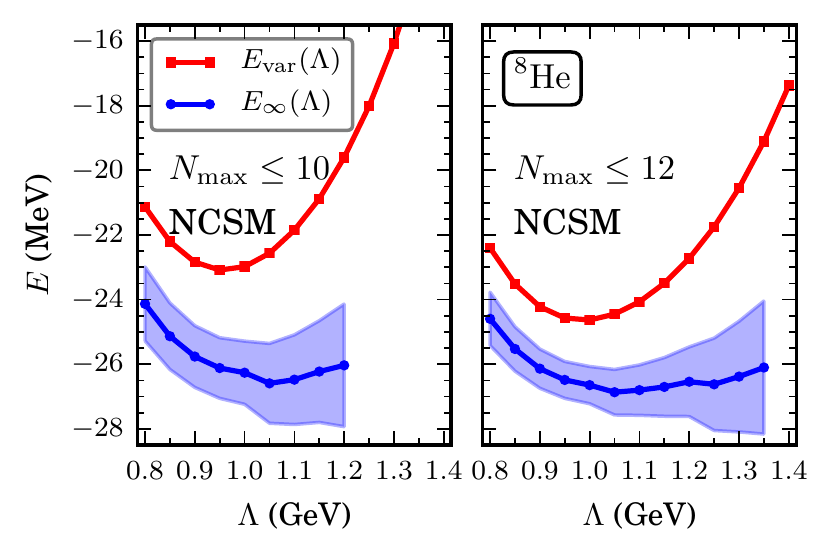} \\
  \includegraphics[width=\columnwidth]{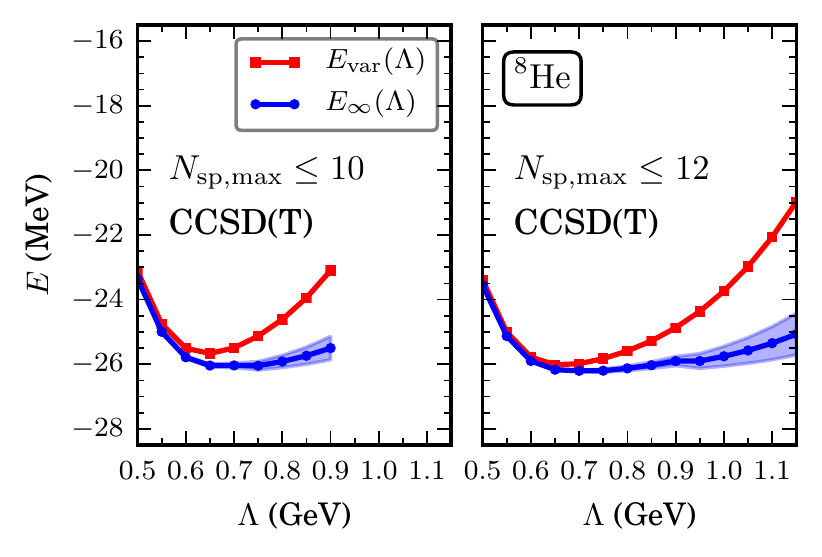}
  \caption{(Color online) Extrapolated energy $E_\infty(\Lambda)$
    (circles) for \nuc{8}{He} from the NCSM (upper panels) and 
    the coupled-cluster method (lower panels) with 
    different model space truncations. The band estimates uncertainties
    from subleading IR corrections. The squares denote the minimum
    energy computed with the respective method as a function of $\Lambda$.
    \label{fig:He8ELO_NCSM_CC}}
\end{figure}

We also performed coupled-cluster
computations~\cite{kuemmel1978,hagen2014} of \nuc{8}{He} in the
Lambda-CCSD(T) approximation~\cite{taube2008}.  These calculations
employ a model space that is a product of single-particle spaces. We
denote the truncation of this space with \nmsp. The relevant IR
lengths [and UV cutoffs via Eqs.~(\ref{L_Lambda})] are taken from
Ref.~\cite{furnstahl2015}.  The results for the ground-state energy
are shown in Fig.~\ref{fig:He8ELO_NCSM_CC}, using the same energy
scale as for the NCSM results.  We note that both methods yield an
extrapolated ground-state energy for \nuc{8}{He} somewhat below
$-26$~MeV. However, the employed model spaces and UV cutoffs $\Lambda$
differ from each other.

For the product space employed in the coupled-cluster computations, we
have not been able to relate the fit parameter $k_\infty$ to an
observable. We note that the result for $k_\infty$ in NCSM and
coupled-cluster calculations differ from each other.

\subsection{\nuc{16}{O}
\label{subsec:A16}}

\begin{figure}[tbp]
  \centering
  \includegraphics[width=\columnwidth]{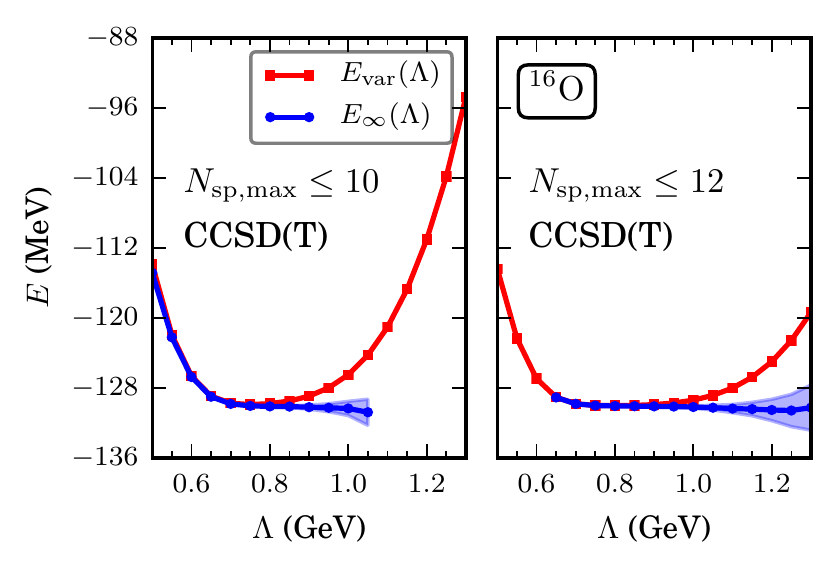}
  \caption{(Color online) Extrapolated energy $E_\infty(\Lambda)$
    (circles) for
    \nuc{16}{O}. The different panels
    correspond to different model space truncations. The 
    bands estimate uncertainties from subleading IR corrections. The
    squares denote the minimum energy computed with the Lambda-CCSD(T) as a
    function of $\Lambda$. 
  \label{fig:O16ELO_CC}}
\end{figure}

\textcite{wendt2015} reported NCSM results for \nuc{16}{O} using the
$NN$ potential NNLO$_{\rm opt}$. These are expensive computations and
we do not repeat them here. In their IR extrapolation, they found a
fit parameter of $k_\infty\approx 0.47$~fm$^{-1}$. This corresponds to
a separation energy of about $S\approx 4.6$~MeV. In \nuc{16}{O}, the
lowest-energetic disintegration threshold is alpha-particle emission,
and the threshold is at $7.16$~MeV experimentally.  The
coupled-cluster computations of \nuc{16}{O} require less effort than
the NCSM and converge rapidly in the model spaces considered in this
work.  The results for the parameters $E_\infty(\Lambda)$ from the
$\chi^2$ fits are shown in Fig.~\ref{fig:O16ELO_CC}.

The extrapolated energy is consistent with the (practically fully
converged) value of $E\approx -130.1$~MeV obtained in model spaces
with $\nmsp=12$. 
We note that the fit of the momentum stabilizes around
$k_\infty\approx 0.97$~fm$^{-1}$, but we have not been able to relate
this value to an observable. We also note that coupled-cluster
calculations in the employed Lambda-CCSD(T)
approximation~\cite{taube2008} would be insensitive to the emission of
alpha particles as this would require at least
four-particle--four-hole excitations. For the employed interaction,
the neutron separation energy is about 20~MeV~\cite{ekstrom2013}.

\section{Summary and discussion}
\label{secsum}
In this paper, we presented three main results. First, we further
advanced the shell-model code \texttt{pANTOINE}, a parallel version of
the Strasbourg shell-model code \texttt{ANTOINE}, to deal with
unprecedented large harmonic oscillator model spaces. This allowed us
to present benchmark results for a variety of light nuclei and to
further explore asymptotic IR extrapolation formulas. The inclusion of
three-nucleon forces into this computational algorithm is an ongoing
task. Second, we performed IR extrapolations of ground-state energies
and radii of $p$-shell nuclei at fixed UV cutoffs in a considerable
range of such cutoffs. This allowed us to improve over previous
extrapolations (taken at very large UV cutoffs) and to present
extrapolations with increased accuracy and precision. In particular,
the extrapolated energies and radii are stable in a range of UV cutoffs
and---for the ground-state energy---consistently improve over the
variational minimum obtained with the available computational
resources. We also found that the momentum $k_\infty$ relevant for the
extrapolation is---within our uncertainties---the same for
ground-state energies and radii.  Third, the extrapolation results
support the hypothesis that the momentum $k_\infty$ (which is the
smallest relevant momentum for ground-state energies and radii)
corresponds to the momentum of the lowest-energetic separation
channel. This identification could be interesting for EFT arguments in
general and for uncertainty quantification in particular. We recall
that it is not really established what is the typical or relevant
momentum scale in finite nuclei. Estimates range from the inverse of
the nucleon-nucleon scattering length on the small side to the Fermi
momentum on the high side. Our results suggest that $k_\infty$ is the
smallest relevant scale. This could imply that the precise
reproduction of nucleon-nucleon scattering data at momenta below
$k_\infty$ is probably not necessary for the computation of well-bound
nuclear states. Of course, (excited) states closer to threshold could
require more accurate properties of the nucleon-nucleon interaction at
lowest momenta.

\begin{acknowledgments}
  We thank Dean Lee for useful discussions. G. H. and T.P. gratefully
  acknowledge the hospitality of the Department of Physics at Chalmers
  during the initial phase of this project. The visit was supported by
  the Swedish Foundation for International Cooperation in Research and
  Higher Education (STINT, Grant No.\ IG2012-5158). This work was also supported
  by the U.S. Department of Energy, Office of Science, Office of
  Nuclear Physics under Awards No.\ DEFG02-96ER40963 (University of
  Tennessee), Nos.\ DE-SC0008499, and No.\ DE-SC0018223 (SciDAC NUCLEI
  Collaboration), the Field Work Proposal ERKBP57 at Oak Ridge
  National Laboratory (ORNL), and under Contract No. DE-AC05-00OR22725
  (Oak Ridge National Laboratory). Some of the computations were
  performed on resources provided by the Swedish National
  Infrastructure for Computing (SNIC) at C3SE (Chalmers) and NSC
  (Link\"oping).
  
\end{acknowledgments}

\appendix

\section{Technical details of the \texttt{pANTOINE} NCSM implementation
\label{sec:pantoine_details}}
The \texttt{pANTOINE} NCSM code uses an iterative scheme to find the
extreme eigenvalues of very large, but relatively sparse, Hermitian
matrices.  Matrix-vector operations consume the most execution time.
Matrix-element indices are computed on-the-fly, as
described in Sec.~\ref{sec:pantoine}. The code runs very efficiently
on single shared-memory machines, although it requires large
memory resources ($\geq 32$ GB).  To handle vectors much larger than
available memory, the operation $y = Mx$ can be split into subsets:
$y_i = (M_{i1}x_1 + M_{i2}x_2 + \dotsb)$.

\begin{figure}[tb]
\begin{center}
\includegraphics[width=\columnwidth]{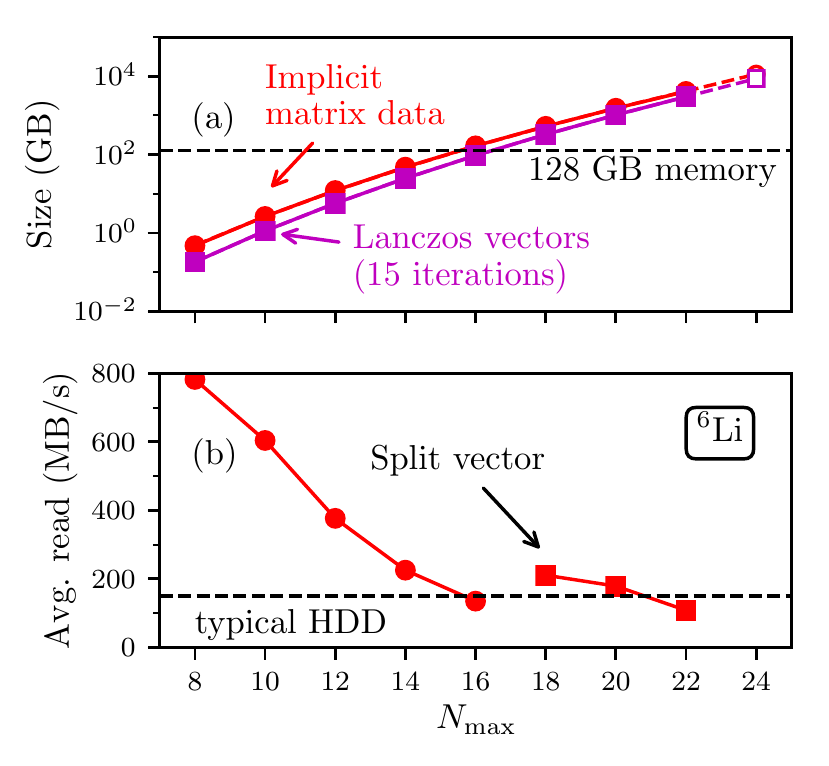}
\end{center}
\caption{\texttt{pANTOINE} scaling plots for the \nuc{6}{Li} nuclear many-body problem as
  a function of the NCSM model space truncation $\nm$. The model space
  dimension is shown in Fig.~\ref{fig:ncsmscaling}. (a) Total storage required for implicit matrix
  construction and for 15 iterations of Lanczos vectors; (b) average
  read speed of implicit matrix data from disk. For $\nm \ge 18$ the
  Lanczos vector is split in several blocks.}
\label{fig:scaling}
\end{figure}

\begin{figure}[tb]
\begin{center}
\includegraphics[width=\columnwidth]{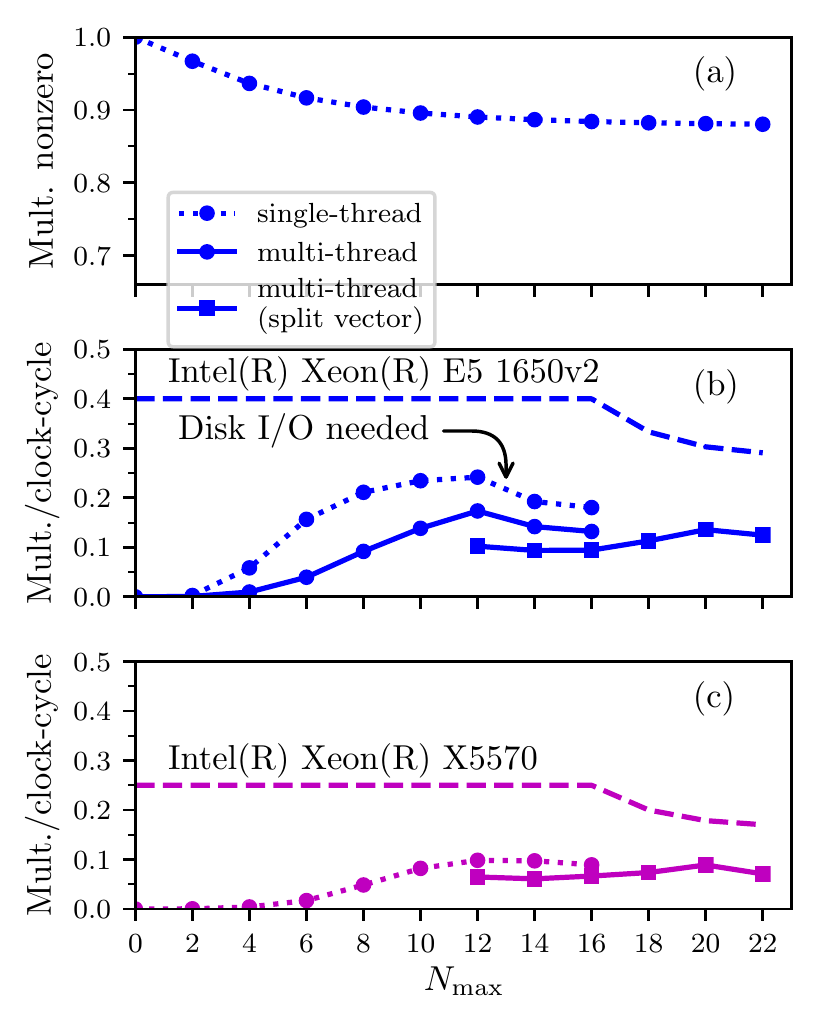}
\end{center}
\caption{\texttt{pANTOINE} scaling plots for the \nuc{6}{Li} nuclear many-body problem
  ($M=1$) as a function of the NCSM model space truncation $\nm$. (a) 
  Multiplication efficiency  (defined as the number of non-zero matrix elements divided by the
  actual number of multiplications being performed); multiplication
  per clock cycle for (b) Intel(R) Xeon(R) E5 1650v2 and (c) Intel(R)
  Xeon(R) X5570 using a single thread (dotted line), multiple threads
  (solid line with circles), and
  multiple-threads with splitting of the Lanczos vector (solid line
  with square symbols).}
\label{fig:memoryeff}
\end{figure}

For two-body nuclear interactions, the code generates the Hamiltonian
matrix on the fly, which removes the need to distribute matrix
elements over thousands of nodes. Accordingly, the results shown in
Figs.~\ref{fig:scaling} and \ref{fig:memoryeff} were
obtained on a single compute node.
The current production version requires node-local disk space for
temporary storage. For job sizes exceeding available memory, it uses
local scratch space efficiently, doing sustained multi-100 MB/s
streaming reads while maintaining close to full multi-core
matrix-vector CPU load.

Still, the most extreme calculations for \nuc{6}{Li} require almost
10~TB of storage for Hamiltonian matrix data and Lanczos vectors
(about half each), see Fig.~\ref{fig:scaling}(a). 
Even though the loaded index triple-values and matrix data is used multiple times
(due to the double loops),
large read speeds, as presented in Fig.~\ref{fig:scaling}(b),
are key for being able to diagonalize matrices with dimensions
surpassing $10^{10}$.
For $\nm \ge 18$ the Lanczos vector is split in several blocks. With a
split vector, mirror blocks are handled separately causing multiple
passes over the same implicit matrix data. Since more data is read in
total, the average read speed increases.

The main improvements of \texttt{pANTOINE} are: (i) extending the
memory management to handle 64-bit offsets and thus allowing much
larger working sets and subset vectors; (ii) multi-threaded inner
loops of the matrix-vector operations using OpenMP; (iii) speeding up
the scratch I/O by the use of raw C-style functions; (iv) asynchronous
read of scratch data, making most I/O be hidden under useful CPU use
(matrix-vector calculations).
Essentially, improvement (i) enables us to handle the larger
model-spaces, while (ii) makes it feasible to run them in reasonable
(but still long) times.  (iii) and (iv) are needed to make (ii)
significant, as I/O-related processing and waiting times otherwise
dominate.


Let us use the $\nm=22$ \nuc{6}{Li} run as a specific example. There
are $5200$ different $nljm$ single-particle states and the resulting
$M=1$ many-body basis has the dimension $2.50 \cdot 10^{10}$.  The
full space corresponds to $\sim200$~GB of storage space per
eigenvector using double-precision for the amplitudes.  Note that the
proton and neutron (three-body) sub-spaces are much smaller than the
full six-body space. The sub-space dimension is $1.83 \cdot 10^{6}$.
There are
$4.88 \cdot 10^{14}$ non-zero matrix elements that are applied on-the-fly
from 4.1~TB of implicit matrix data.
However, since the data is used in different combinations,
the program actually reads 24.9~TB of data from disk per
iteration. At every iteration, the full matrix-vector operation
requires $5.54 \cdot 10^{14}$ multiplications, which takes a couple of
days on a single compute node.
The difference between number of non-zero elements and total number of
multiplications come from the use of precalculated index-triples.  It
is not possible to avoid matrix-elements
that vanish due to non-trivial cancellations of Clebsch-Gordan coefficients.
Fig.~\ref{fig:memoryeff}(a) shows how this inherent inefficiency develops
with model space size.

In order for I/O reads not to act as a significant bottle-neck, a fast
scratch storage component of the computer is required.  Due to the
multi-threaded calculations needing separate large output vectors, it
is also beneficial to use fewer but fast and efficient processor
cores.  For this calculation we used a purpose-built machine with a
6-core Xeon E5-1650v2 CPU, 128~GB of RAM and $10 \times 4$~TB HGST NAS
disks in a RAID 5 configuration, capable of streaming scratch data at
$\sim1$~GB/s.  In this case, streaming reads averaging 108~MB/sec, see
Fig.~\ref{fig:scaling}(b), were done in parallel with maintaining a
very large CPU load.
%
The decrease in multiplication efficiency shown in
Fig.~\ref{fig:memoryeff}(b) at $\nm \approx 12-14$ is due to
operating-system disk cache
space running out, necessitating I/O each iteration.  The low
efficiency at the smallest $\nm$ is due to block-scheduling
overhead each iteration, which also is included in the measurements.
The multiplication efficiencies for multi-thread runs with split vectors
were measured with the Lanczos vectors divided in four roughly equal
pieces, except for $\nm=22$ where sixteen pieces were used.

Despite the heavy I/O and on-the-fly computation of matrix element
indices the code performed $2.40 \cdot 10^{9}$ multiplications/sec in
average. With $6 \times 3.6$~GHz
available, this implies an impressive load of 0.111
multiplications per clock cycle and core, see
Fig.~\ref{fig:memoryeff}(b).
The dashed curves show the upper efficiency limit given by the
available execution resources of each processor type, considering the
assembler code of the dominating computational kernel.
%

\section{\nuc{3}{He}
\label{sec:He3}}

\begin{figure}[htbp]
  \centering
  \includegraphics[width=\columnwidth]{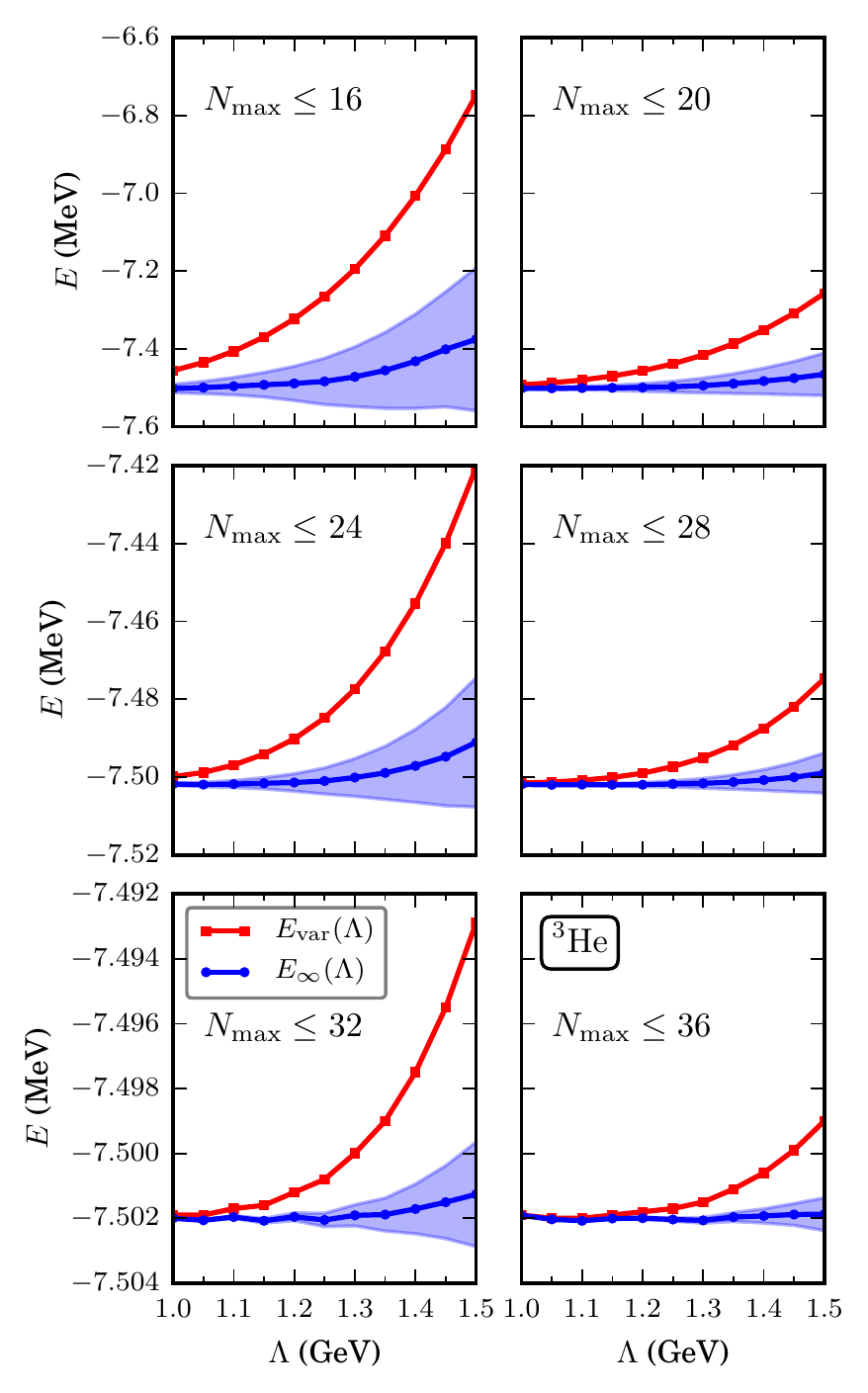}
  \caption{(Color online) Extrapolated energy $E_\infty(\Lambda)$
    (circles) for
    \nuc{3}{He}. 
    See caption of Fig. \ref{fig:H3E0LO} for further details.
  }
  \label{fig:He3E0LO}
\end{figure}

\begin{figure}[htbp]
  \centering
  \includegraphics[width=\columnwidth]{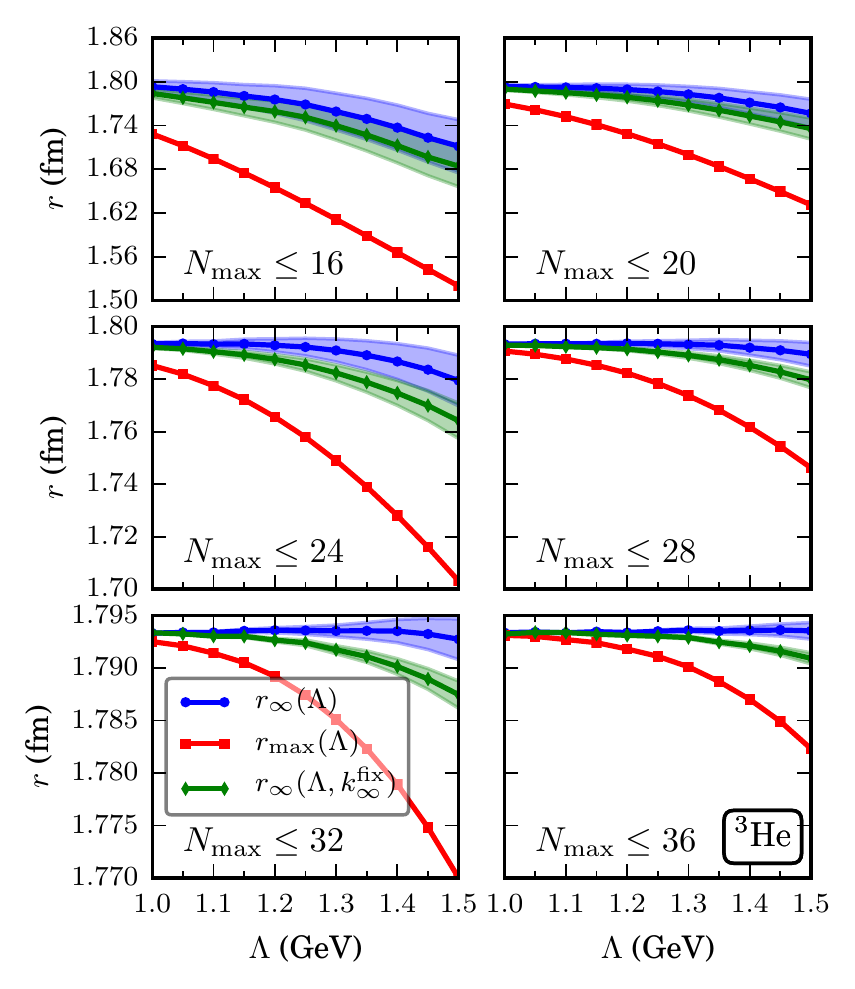}
  \caption{(Color online) Extrapolated ground-state (point-proton)
    radii $r_\infty(\Lambda)$
    (circles) for
    \nuc{3}{He}. The different panels
    correspond to different NCSM model space truncations from
    $\max(N_\mathrm{max})=16$ to $\max(N_\mathrm{max})=36$. 
    See caption of Fig. \ref{fig:H3RLO} for further details.
  }
  \label{fig:He3RLO}
\end{figure}

While the extrapolation results for \nuc{3}{H} were shown already in
Sec.~\ref{subsec:A3}, the corresponding results for \nuc{3}{He} energies
(radii) are shown in Fig.~\ref{fig:He3E0LO} (Fig.~\ref{fig:He3RLO}).

\begin{figure}[htbp]
  \centering
  \includegraphics[width=\columnwidth]{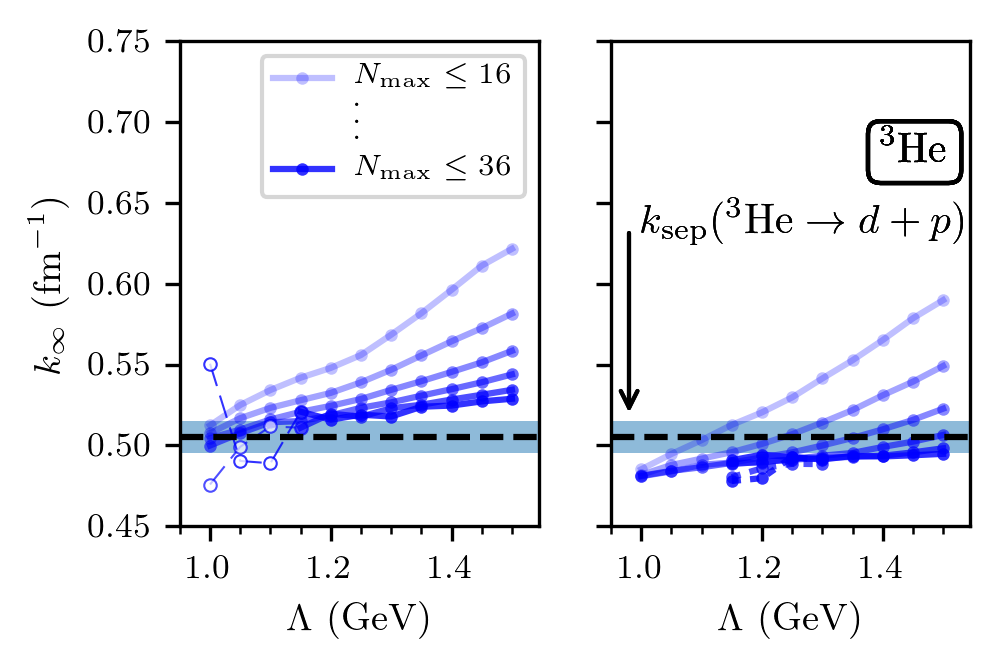}
  \caption{(Color online) Fit parameter $k_\infty(\Lambda)$ for
    \nuc{3}{He} energy extrapolation (left panel) and radius
    extrapolation (right panel) for different NCSM model space
    truncations from $\max(N_\mathrm{max})=16$ to
    $\max(N_\mathrm{max})=36$. 
    Open symbols denote results for which UV corrections are expected
    to be larger than IR ones, and the corresponding fits are unreliable.
    The lowest, theoretical separation
    momentum is given as a dashed line with an uncertainty band.}
  \label{fig:He3kLO}
\end{figure}

The values for $k_\infty$ resulting from the fit of Eq.~(\ref{master})
are shown in Fig.~\ref{fig:He3kLO} and we find that a stable region is
reached for large enough UV scales. The value in this stable region
agrees very well with the \nuc{3}{He} separation momentum for
$\nuc{3}{He}\to d+p$, which is $k_\infty\approx 0.50$~fm$^{-1}$. We
note that this momentum is not well separated from the momentum
$k_\mathrm{sep}(\nuc{3}{He} \to p+p+n) \approx 0.60$~fm$^{-1}$ for three-body breakup.

\begin{figure}[htbp]
  \centering
  \includegraphics[width=\columnwidth]{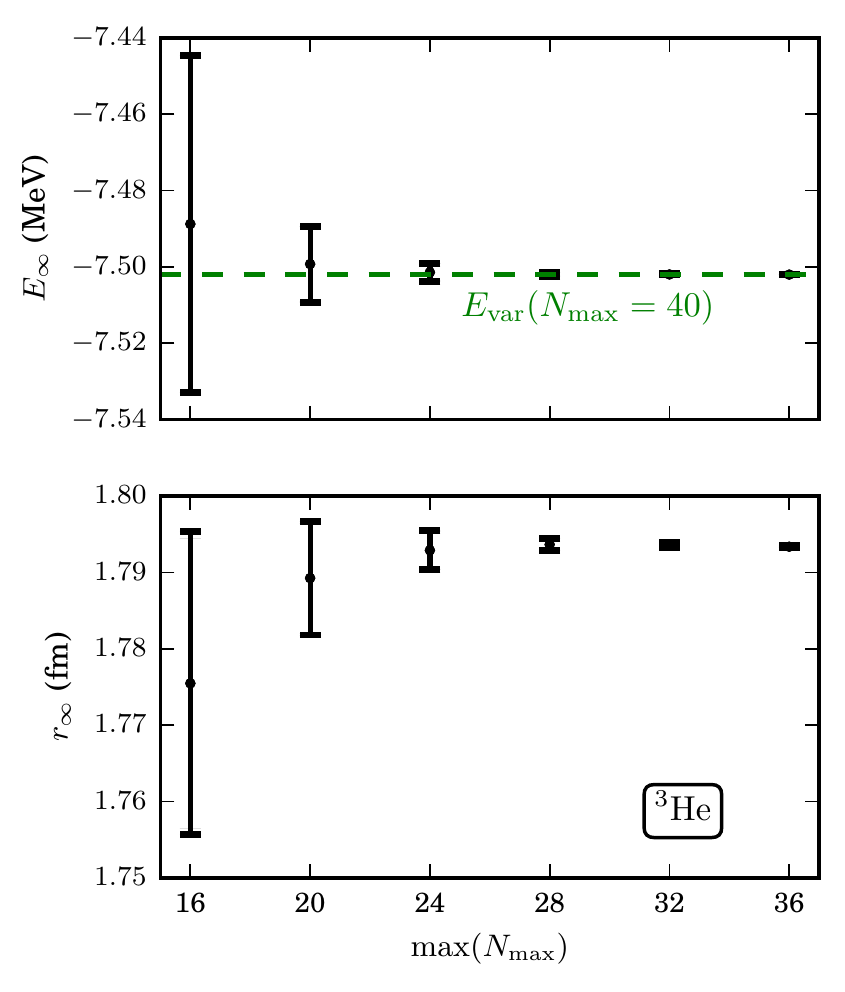}
  \caption{(Color online) Recommended results for the
    \nuc{3}{He} energy (upper panel) and radius
     (lower panel) for different NCSM model space
    truncations from $\max(N_\mathrm{max})=16$ to
    $\max(N_\mathrm{max})=36$.}
  \label{fig:He3recLO}
\end{figure}

Based on these observation we use the extrapolations at (fixed)
$\Lambda=1200$~MeV to extract a sequence of recommended results for
the ground-state energy and the point-proton radius as a function of
the model-space truncation; see Fig.~\ref{fig:He3recLO}.  Overall, the
quality of the results for \nuc{3}{He} is similar to those obtained for
\nuc{3}{H} in Sect.~\ref{subsec:A3}.

\section{Asymptotic normalization coefficient}
\label{App_ANC}
We want relate the parameter $a_0$ of Eq.~(\ref{master}) to the
asymptotic normalization coefficient. For the two-body system, the relation
\be
\label{anc-2body}
a_0={\hbar^2 k_\infty \gamma_\infty^2\over \mu}
\ee
was derived in Ref.~\cite{furnstahl2014} using scattering
theory. Here, $\mu$ is the reduced mass, $k_\infty$ the canonical
momentum corresponding to the two-particle distance
$|\vec{r}_1-\vec{r}_2|$, and $\gamma_\infty$ the corresponding ANC.
It is straightforward to derive the analogous expression for any
two-body breakup. In the orthogonal Jacobi coordinates~(\ref{jacobi}),
this yields
\be
\label{anc}
a_0={\hbar^2 k_\infty \gamma_\infty^2\over m} .  
\ee
For completeness, we also give a derivation of this result using the
Hamiltonian. The derivation adapts the approaches of
\citeauthor{luscher1985}~\cite{luscher1985} and
\citeauthor{koenig2012}~\cite{koenig2012} to our case.

Let $\rho_1$ be the orthogonal Jacobi coordinate that describes the
separation between two clusters, and $m$ the nucleon mass. Then the
bound-state wave function of the two clusters is $\psi_{\rm sep}(\rho_1) =
u_{\rm sep}(\rho_1)/\rho_1$, the separation energy is $E_{\rm sep} = - \hbar^2
k_{\rm sep}^2/(2m)$, and for interactions of range $R$, we have for $\rho_1\gg
R$
\be
u_{\rm sep}(\rho_1)=\gamma_{\rm sep} e^{-k_{\rm sep} \rho_1} .
\ee
Here the ANC ensures that the wave function is properly
normalized. Let us consider the wave function
\ba
u(\rho_1) = \left\{\begin{array}{ll}
\alpha u_{\rm sep}(\rho_1) , & \rho_1<R,\\
\alpha \gamma_{\rm sep} e^{-k_{\rm sep} R}\frac{e^{-k_{\rm sep} \rho_1} -e^{-k_{\rm sep} (2L-\rho_1)}}{e^{-k_{\rm sep} R} -e^{-k_{\rm sep} (2L-R)}}, & R\le \rho_1\le L. \end{array}\right. \nonumber
\ea
Here, $u(\rho_1)$ is the exact finite-space wave function with
separation energy $E_{\rm sep}$ of the Hamiltonian $H$ for
$R\le \rho_1 \le L$. We have $u(L)=0$, since $L$ is our hard-wall
radius, and we assume $R\ll L$. Inspection shows that the normalization
constant $\alpha = 1 +{\cal O}(k_{\rm sep} L e^{-2k_{\rm sep} L})$. We
also have
\ba
u(R) &=& \alpha\gamma_{\rm sep} e^{-k_{\rm sep} R} , \nonumber\\
u'(R+\varepsilon) &=& -k_{\rm sep} u(R)\coth \left( k_{\rm sep} \left[
    L-R \right] \right) ,   \nonumber\\
u'(R-\varepsilon) &=& -k_{\rm sep} u(R) ,
\ea
where $\varepsilon \ll R$ is a small distance.
We see that the wave function is continuous at $\rho_1=R$, but its
derivative makes a jump
\be
u'(R+\varepsilon) - u'(R-\varepsilon) \approx -2k_{\rm sep} e^{-2k_{\rm sep} (L-R)} u(R)  
\ee
at $\rho_1=R$. Thus, $u(\rho_1)$ is an exact finite-space
eigenfunction with energy $E_{\rm sep}$ of the Hamiltonian $H'=H+V$, with
\be
V(\rho_1) = -{\hbar^2 k_{\rm sep}\over m} e^{-2k_{\rm sep} (L-R)} \delta(\rho_1-R) . 
\ee
We note that $V$ is exponentially small. Thus, $u(\rho_1)$ is an
exponentially good approximation of the eigenfunction of $H$ in a
finite space. Let $u_L(\rho_1)$ denote the exact finite-space eigenstate of
$H$, with eigenvalue $E_L$. We have
\be
u_L(\rho_1) = \beta u(\rho_1) + \delta u(\rho_1) , 
\ee
with $\beta=1+{\cal O}(e^{-k_{\rm sep} L})$ and $\delta u(\rho_1)= {\cal
  O}(e^{-k_{\rm sep} L})$, and $\langle u|\delta u\rangle =0$. Thus
\ba
\langle u_L |H|u\rangle &=& \langle u_L |(H'-V)|u\rangle \nonumber\\
&=& E_{\rm sep} \langle u_L|u\rangle +{\hbar^2 k_{\rm sep}\over m} e^{-2k_{\rm sep} L} u_L(R)u(R) , \nonumber\\
\langle u_L |H|u\rangle &=& E_L \langle u_L|u\rangle ,
\ea
from acting with $H$ to the right and to the left, respectively. As
$\langle u_L|u\rangle=1+{\cal O}(e^{-k_{\rm sep} L})$ and $u(R)\approx
u_L(R)\approx u_{\rm sep}(R)$ up to exponentially small corrections, we
get
\be
\label{result}
E_L- E_{\rm sep} = {\hbar^2k_{\rm sep}\gamma_{\rm sep}^2\over m}e^{-2k_{\rm sep} L}
\ee
in leading order. This is the desired result.

The generalization to many-body bound states and two-cluster breakup
is straightforward, e.g., by following
\citeauthor{koenig2018}~\cite{koenig2018}. In this case, the $A$-body
wave function is the product
\ba
\lefteqn{\Psi_A(\vec{r}_1,\ldots,\vec{r}_A) = }\nonumber\\
  &&\equiv \Psi_a(\vec{r}_1,\ldots,\vec{r}_a)
\Psi_{A-a}(\vec{r}_{a+1},\ldots,\vec{r}_A) \psi_{\rm sep}(\vec{\rho}_1) .
\ea
Here, $\rho_1$ denotes the orthogonal Jacobi coordinate between the
clusters of $a$ and $A-a$ particles, respectively. For ease of
notation we suppressed the spin/isospin degrees of freedom, and it is
also understood that the overall wave function $\Psi_A$ needs to be
properly antisymmetrized. The separation momentum is
\be
\label{ksep}
k_{\rm sep} = \hbar^{-1}\sqrt{2m\left(B_A-B_a-B_{A-a}\right)} , 
\ee
and $B_n$ is the binding energy of the cluster with mass number
$n$. We can now follow the derivation of Ref.~\cite{koenig2018} and
arrive at the result~(\ref{result}) for the correction to the
separation energy.

In contrast to Ref.~\cite{koenig2018}, the nucleon mass $m$ (and not a
reduced mass) enters the expression~(\ref{ksep}), because we employ an
orthogonal Jacobi coordinate $\rho_1$ instead of the physical
separation,
\be
\vec{r}\equiv {1\over a}\sum_{i=1}^a \vec{r}_i - {1\over  A-a}\sum_{i=a+1}^A \vec{r}_i,
\ee
of the center of masses between both clusters. Thus, our asymptotic
normalization coefficient $\gamma_{\rm sep}$ needs to be rescaled
before it can be compared to data. We have $\rho_1 = \sqrt{a(A-a)/A}
|\vec{r}|$. Thus our $k_{\rm sep}$ is the physical separation momentum
times the factor $\sqrt{A/[a(A-a)]}$, and our ANC is the
physical ANC times $(A/[a(A-a)])^{1/4}$.

\bibliography{thomas,refs,ref2}

\end{document}